\documentclass[pdflatex,sn-nature]{sn-jnl}% Style for submissions to Nature Portfolio journals (Scientific Reports)
\graphicspath{{Images/}{images/}{figures/}{./}}% figure folder
\usepackage{hyperref}
\usepackage{etoolbox}

\usepackage{tikz}%
\definecolor{orcidgreen}{HTML}{A6CE39}%
\renewcommand{\orcidlogo}{\raisebox{-0.15ex}{\tikz[baseline=-0.5ex]{%
  \fill[orcidgreen] (0,0) circle (0.9ex);%
  \node[text=white,font=\fontsize{4.5}{4.5}\selectfont\bfseries\sffamily,inner sep=0pt] at (0,0) {iD};}}}%
\usepackage{graphicx}%
\usepackage{multirow}%
\usepackage{amsmath,amssymb,amsfonts}%
\usepackage{amsthm}%
\usepackage{rsfso}%
\usepackage[title]{appendix}%
\usepackage{xcolor}%
\usepackage{textcomp}%
\usepackage{manyfoot}%
\usepackage{etoolbox}% for \pretocmd
\usepackage{etoolbox}% for \pretocmd
\usepackage{etoolbox}%
\usepackage{etoolbox}
\usepackage{etoolbox}% for \pretocmd
\usepackage{etoolbox}
\usepackage{booktabs}%
\usepackage{algorithm}%
\usepackage{algorithmicx}%
\usepackage{algpseudocode}%
\usepackage{listings}%
\theoremstyle{thmstyleone}%
\theoremstyle{thmstyletwo}%

\theoremstyle{thmstylethree}%

\unnumbered% unnumbered headings, as used by Scientific Reports

\begin{document}

\title[Magnonic cavity-enhanced NV magnetometry]{Magnonic cavity-enhanced quantum magnetometry with nitrogen-vacancy centre ensembles}

\author*[1]{\fnm{Nikhil} \sur{Kumar}\,\orcid{https://orcid.org/0000-0003-3213-505X}}\email{nikhilkumarcs@nitc.ac.in}
\affil*[1]{\orgdiv{Department of Electronics and Communication Engineering}, \orgname{National Institute of Technology Calicut}, \orgaddress{\city{Kozhikode}, \postcode{673601}, \state{Kerala}, \country{India}}}

\abstract{\unboldmath Nitrogen-vacancy (NV) centre ensembles are leading room-temperature magnetometers, but their sensitivity is usually extracted from a single resonance and relies on external microwave and bias-field hardware. This work proposes a
chip-scale NV magnetometer built around a one-dimensional YIG/CoFeB magnonic
crystal, in which a single structural defect forms a passive magnonic cavity and
integrated permanent magnets supply the bias field. Micromagnetic simulations show
a defect-localised spin-wave mode at 2.910 GHz inside the 2.5--3.0 GHz magnonic
band gap, 40 MHz from the NV zero-field splitting, with a band structure that
matches an independent plane-wave calculation. A Tavis--Cummings/Lindblad model in
which the microwave reaches the NV ensemble through this mode predicts a
hyperfine-resolved optically detected magnetic resonance spectrum that is strong
only within the mode linewidth, and a dispersive homodyne response with one steep
edge per $^{14}$N hyperfine line. The master-equation solutions agree with the
closed-form homodyne steady state ($R^{2}>0.9999$) and with an approximate analytic
ODMR theory (peak positions within 0.012 MHz). For a critically coupled readout with
room-temperature electronics, combining the three edges gives a projected
sensitivity of 93 pT$/\sqrt{\mathrm{Hz}}$, a $\sqrt{3}$-fold gain over the best
single edge. Mapping the sensitivity against the NV--magnon coupling reveals an
optimum near $g_{\mathrm{ens}}/2\pi\approx0.2$--0.4 MHz, where it reaches
72 pT$/\sqrt{\mathrm{Hz}}$ for a 0.64 MHz spin linewidth and 19 pT$/\sqrt{\mathrm{Hz}}$
for 0.17 MHz, with thermal limits of 6.6 and 1.7 pT$/\sqrt{\mathrm{Hz}}$. The
antenna--mode coupling, the NV--magnon coupling and the spin linewidth are
therefore the main design levers.}


\maketitle

%% =====================================================================
\section{Introduction}

The ensembles of nitrogen-vacancy (NV) centres in diamond are among the most widely used solid-state platforms for room-temperature magnetometry, owing to their optical addressability and long spin coherence times\cite{taylor2008high,doherty2013nitrogen,rondin2014magnetometry,schirhagl2014nitrogen,barry2020sensitivity}. Researchers have developed single NV spins and NV ensembles into a quantum sensing toolkit for condensed-matter imaging, biology, and geoscience, \cite{degen2017quantum,casola2018probing}, and naval magnetic anomaly detection (MAD) has emerged as one of the most demanding applications.\cite{sheinker2009magnetic}

The photon-collection efficiency of the optical readout channel fundamentally limits the ensemble sensitivity. Coupling the ensemble to a microwave cavity addresses this limitation by transducing the spin signal through a broadband microwave channel,\cite{eisenach2021cavity,wang2024spin} building on coherent magnon--photon coupling in cavity magnonics.\cite{huebl2013high,zhang2014strongly,tabuchi2014hybridizing} In a strongly driven, weakly coupled regime, H.~Kumar \emph{et al.}\cite{kumar2026cavity} recently showed that such a cavity creates doubly dressed NV states: Autler--Townes-split\cite{autler1955stark} hyperfine transitions are further split by vacuum-Rabi coupling to the cavity, producing nine spectrally resolved peaks. Together, these peaks provide nearly a threefold sensitivity enhancement over single-peak readout, with a measured sensitivity of 12 pT$/\sqrt{\mathrm{Hz}}$ and a projected path of 100 fT$/\sqrt{\mathrm{Hz}}$.

NV spins also couple directly to magnons in an adjacent ferromagnetic film through the dipolar interaction, so that a magnon mode can act as a coherent near-resonant drive instead of a microwave antenna\cite{andrich2017long,bertelli2020magnetic,fukami2021opportunities}. Magnonic crystals, the magnetic counterpart of photonic crystals, are periodically modulated magnetic media in which Bragg scattering opens spin-wave band gaps\cite{vasseur1996magnon,nikitov2001spin}. Yttrium iron garnet (YIG) supports low-loss magnonic filters, delay lines, and field sensors\cite{kruglyak2010magnonics,krawczyk2014review,chumak2015magnon,chumak2017magnonic,barman2021magnonics,inoue2011investigating}. Most importantly for NV dressing, a local break in periodicity pulls a spatially localised high-quality factor mode into the band gap, in direct analogy with a photonic-crystal defect cavity\cite{joannopoulos2008photonic,kruglyak2006spin,gallardo2018symmetry}. Earlier studies, including work by the author, have demonstrated such defects and cavity modes in bicomponent stripe and garnet-based structures closely related to the one studied here\cite{kumar2013spin,kumar2018resonant,kumar2024exciting,santos2023magnon,qin2021nanoscale,ordonezromero2025spin}.

A magnonic cavity offers a practical advantage over the actively pumped, resonantly matched dielectric resonator of the cavity-driven scheme, which requires continuous drive power, precise tuning, and dedicated microwave hardware. A simple antenna or stripline patterned on the magnetic film can excite the magnonic cavity, so a magnon-mediated multispectral magnetometer could take the form of a compact chip-scale sensing element. This work numerically investigates whether a passive one-dimensional (1D) YIG/CoFeB magnonic cavity can reproduce the multispectral dressing physics of ref.~\cite{kumar2026cavity}. Micromagnetic simulations with MuMax3\cite{vansteenkiste2014design} identify the defect-localised mode and its dynamic dipolar field at the NV site; a plane-wave method (PWM)\cite{krawczyk2008plane,klos2013magnonic,kumar2013spin,venkat2014micromagnetic} independently verifies the band structure; and a Tavis--Cummings/Lindblad model solved in QuTiP\cite{johansson2012qutip} converts this field into hyperfine-resolved spectra and multispectral sensitivity.

%% =====================================================================
\section{Results}

\subsection{Theoretical framework}

A Tavis--Cummings-type Hamiltonian\cite{tavis1968exact} describes an NV ensemble coupled to a single bosonic mode $a$\cite{kumar2026cavity}. Because the microwave field does not drive transitions between hyperfine sublevels of $^{14}$N, each nuclear spin state $m_{I}\in\{-1,0,+1\}$ forms an independent two-level system on the $m_{s}=0\leftrightarrow m_{s}=+1$ transition. A strong classical field splits each transition into two dressed states separated by the Rabi frequency $\Omega$, giving a Mollow triplet\cite{mollow1969power} per hyperfine state and nine lines in total; coupling to the bosonic mode then splits these lines further. In a frame rotating at the drive frequency $\omega_{d}$ (with $\hslash=1$),

\begin{equation}
H=\Delta_{c}\,a^{\dagger}a+\sum_{j=-1}^{+1}\left[\frac{\Delta_{j}}{2}\,\sigma_{j}^{z}+g_{\mathrm{ens}}\left(a\,\sigma_{j}^{+}+a^{\dagger}\sigma_{j}^{-}\right)+\frac{\Omega}{2}\left(\sigma_{j}^{+}+\sigma_{j}^{-}\right)\right],\label{eq:hamiltonian}
\end{equation}

\noindent where $\Delta_{c}=\omega_{c}-\omega_{d}$ and $\Delta_{j}=\omega_{j}-\omega_{d}$ are the re-tunings of the mode and spin, $\sigma_{j}^{\pm}$ and $\sigma_{j}^{z}$ are the spin operators of sector $j$, and $g_{\mathrm{ens}}=g\sqrt{N_{s}}$ is the collective coupling of the spins $N_{s}$. In the present work, $a$ is the defect-localised magnon mode rather than a cavity photon. The transverse dipolar field of this mode at the NV site sets the Rabi frequency $\Omega$, and the field per magnon sets $g_{\mathrm{ens}}$. Both architectures therefore share the same Hamiltonian; they differ in the mode linewidth, in the coupling of the mode to the external microwave line and in the coupling strengths, which together set the sensitivity. The Lindblad master equation\cite{breuer2002theory} accounts for dissipation,
\begin{equation}
\dot{\rho}=-i\left[H,\rho\right]-\frac{\gamma_{0}}{2}\,\mathcal{D}[a]\rho-\sum_{j,k}\frac{\gamma_{k}}{2}\,\mathcal{D}\!\left[\sigma_{j}^{k}\right]\rho,\label{eq:lindblad}
\end{equation}
with $\mathcal{D}[x]\rho=\{x^{\dagger}x,\rho\}-2x\rho x^{\dagger}$, $\sigma_{j}^{k}\in\{\sigma_{j}^{-},\sigma_{j}^{z}\}$, mode loss rate $\gamma_{0}$, spin relaxation rate $\gamma_{\parallel}$ and dephasing rate $\gamma_{\perp}$.

\subsection{Magnonic band gap and the defect-localised mode}

\begin{figure}[htbp]
\centering
\includegraphics[width=\textwidth]{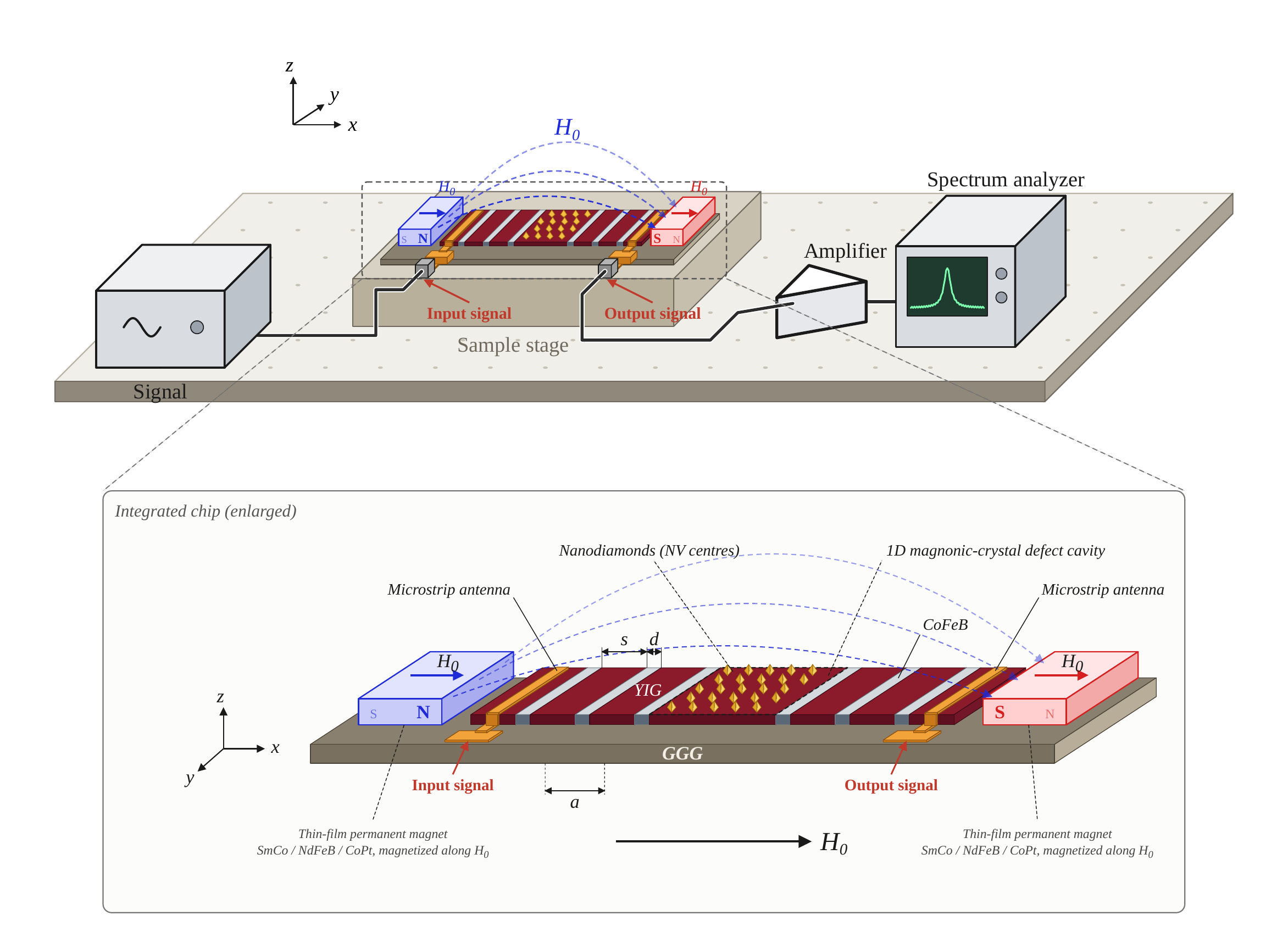}
\caption{\textbf{Schematic of the proposed on-chip magnonic-crystal/NV-centre device and its measurement setup.} Top: the integrated chip is mounted on a sample stage. A microwave signal generator feeds the input microstrip antenna through an SMA connector (input signal). Spin waves launched by the antenna propagate along $x$ through the magnonic crystal; the output microstrip antenna detects them inductively (output signal), and the spectrum analyser amplifies and records the transmission spectrum. No external electromagnet is required. Bottom (enlarged): the device is a one-dimensional magnonic crystal formed by YIG segments of length $s$ separated by CoFeB stripes of width $d$ (lattice constant $a=s+d$) on a GGG substrate. A longer central YIG segment acts as a defect cavity that confines a spin-wave mode within the magnonic band gap, and nanodiamonds containing NV centres are deposited on this cavity. Thin-film permanent magnets (for example SmCo, NdFeB or CoPt), patterned on the same substrate at both ends of the crystal and magnetised along $H_{0}$, present N and S poles to the crystal; their stray field (dashed lines) supplies the in-plane bias field $H_{0}\parallel x$, parallel to the direction of periodicity. The input and output microstrip antennas span the first and last YIG segments, run parallel to $y$, and are fed from pads at the substrate edge. Drawing not to scale.}\label{fig:device}
\end{figure}

\begin{figure}[htbp]
\centering
\includegraphics[width=0.85\textwidth]{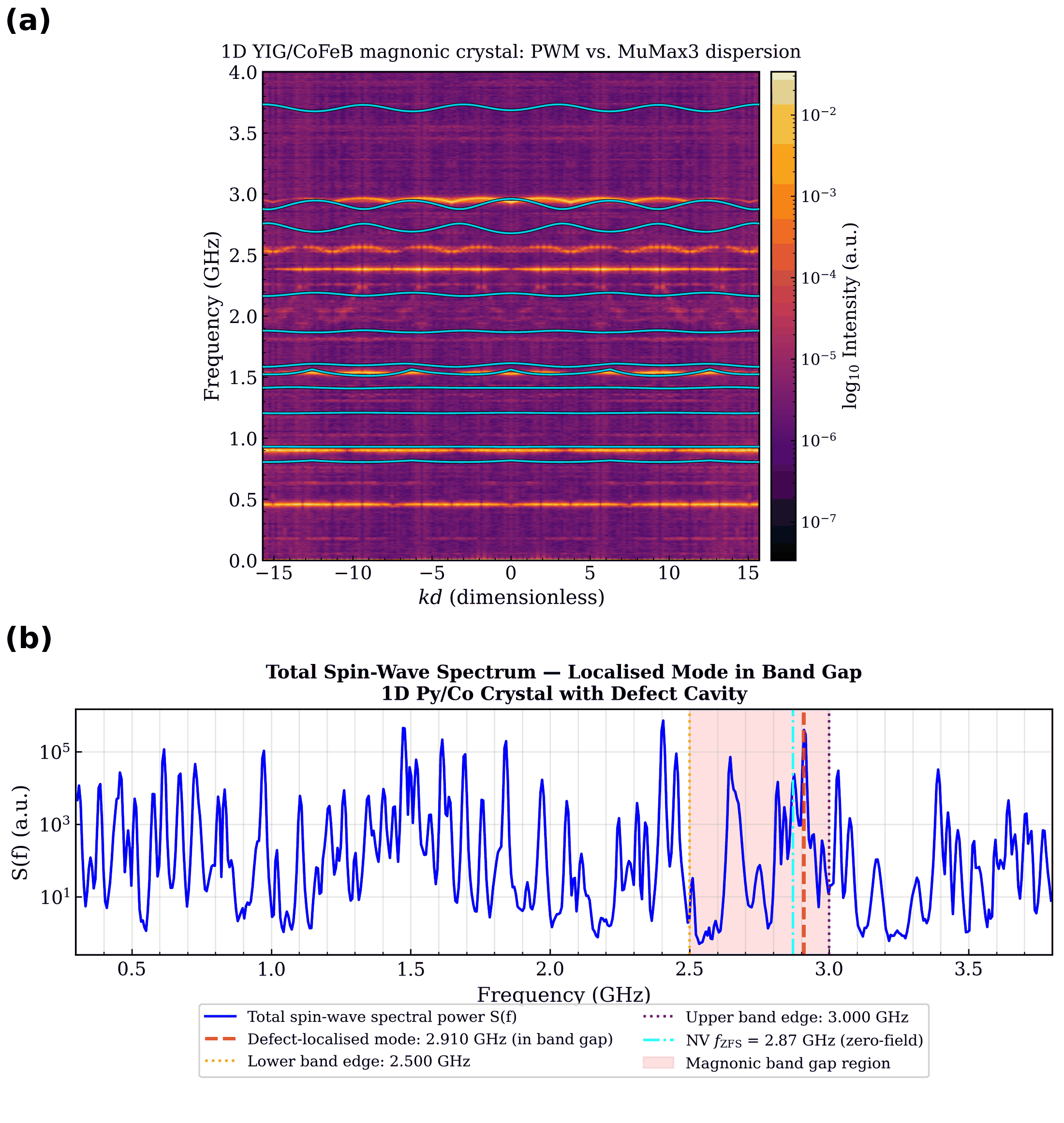}
\caption{\textbf{Band structure and the defect-localised mode.} (a) Micromagnetically simulated dispersion of the pristine 1D YIG/CoFeB crystal (colour map, $\log_{10}$ intensity of the space--time Fourier transform of the ring-down response versus normalised Bloch wave vector), overlaid without adjustable parameters on the independently computed plane-wave bands (cyan curves). (b) Total spin-wave spectral power $S(f)$ of the crystal with the defect, showing the magnonic band gap (shaded, 2.5--3.0 GHz) and the single defect-localised mode at 2.910 GHz (red dashed line) just above the NV zero-field splitting at 2.87 GHz (cyan dash-dotted line).}\label{fig:bandgap}
\end{figure}

Fig. ~\ref{fig:device} shows the proposed device and its measurement setup. The chip carries a 1D magnonic crystal of YIG segments separated by CoFeB stripes on a gadolinium gallium garnet (GGG) substrate. A longer central YIG segment forms a defect cavity at $x=16{,}250$ nm, and NV-containing nanodiamonds sit on this cavity. Thin-film permanent magnets patterned at both ends of the crystal supply the in-plane bias field $H_{0}\parallel x$, so no external electromagnet is needed, and two microstrip antennas across the first and last YIG segments launch and detect spin waves for transmission spectroscopy. The simulations represent the magnet stray field by a uniform bias $B_{\mathrm{ext}}=101.4$ mT along $x$, under which the MuMax3 relax() routine brings the crystal to equilibrium (static stray field maps in Supplementary Figs.~S1 and S2). Two independent calculations then establish the band structure of the pristine crystal (Fig.~\ref{fig:bandgap}a). The micromagnetic dispersion and the PWM bands agree without adjustable parameters: the PWM reproduces the dispersive doublets near 0.85--0.95 and 2.7--3.0 GHz and the zone-folded band near 1.5 GHz with its Bragg-like kinks, while a few weak features absent from the linear PWM point to higher-order spin-wave processes (Supplementary Note~2).

\begin{figure}[htbp]
\centering
\includegraphics[width=\textwidth]{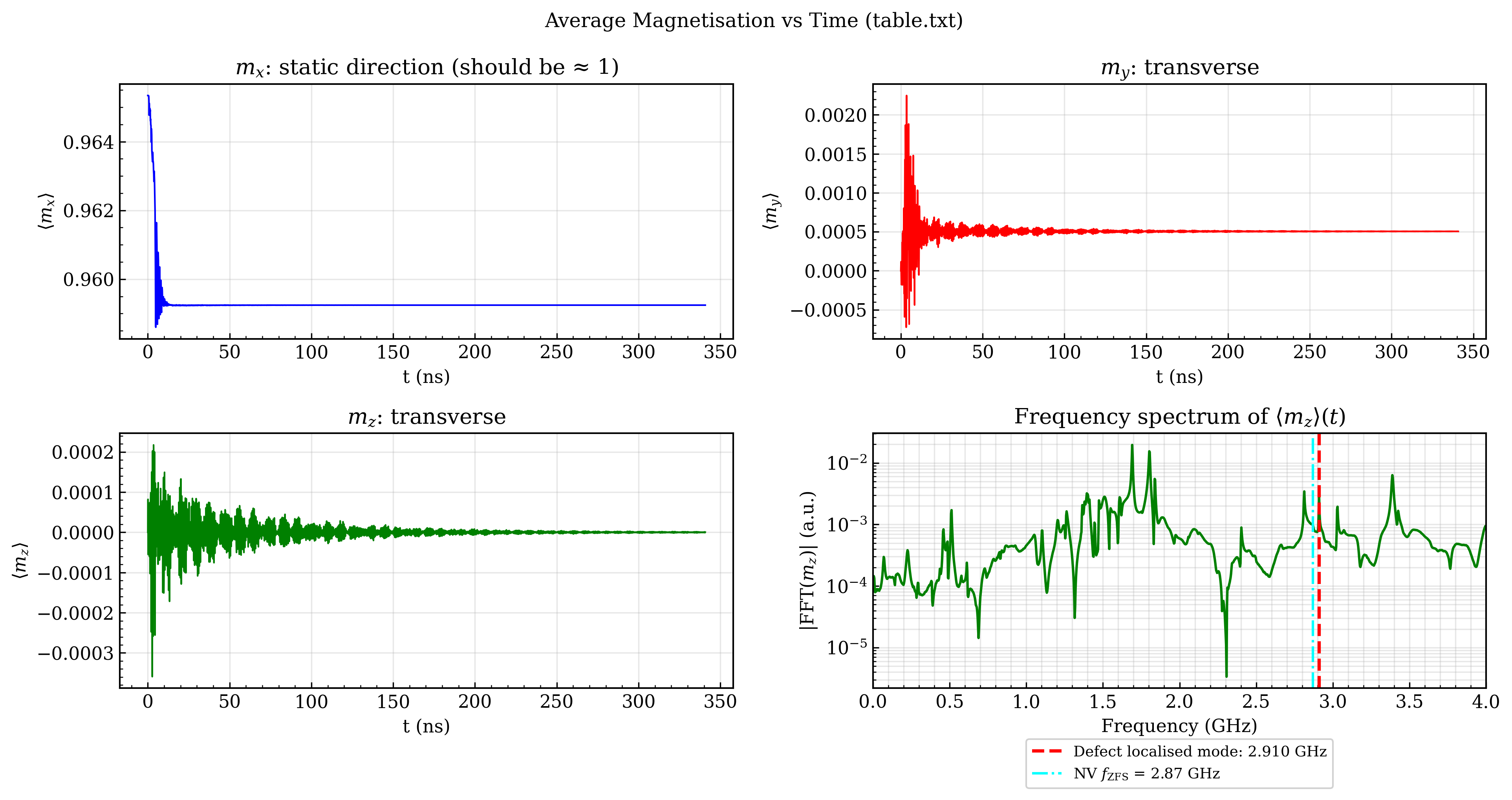}
\caption{\textbf{Magnetisation dynamics after broadband excitation.} Volume-averaged magnetization components $\langle m_{x}\rangle$, $\langle m_{y}\rangle$ and $\langle m_{z}\rangle$ versus time after the broadband sinc pulse applied uniformly across the device. Bottom right: Fourier spectrum of $\langle m_{z}(t)\rangle$, showing the defect-localised resonance at 2.910 GHz (red dashed line) close to the NV zero-field splitting at 2.87 GHz (cyan dash-dotted line).}\label{fig:time-traces}
\end{figure}

A spatially uniform broadband sinc-pulse field (10 mT amplitude, 5 GHz cut-off) then excites the relaxed crystal, and the simulation records the volume-averaged magnetisation (Fig.~\ref{fig:time-traces}). The static component $\langle m_{x}\rangle$ settles within about 25 ns, while the transverse components $\langle m_{y}\rangle$ and $\langle m_{z}\rangle$ decay over roughly 150 ns as the excited spin-wave modes dephase and damp. The Fourier transform of $\langle m_{z}(t)\rangle$ gives the spin-wave spectral power $S(f)$ (Fig.~\ref{fig:bandgap}b), which shows an allowed-band continuum and a band gap between 2.5 and 3.0 GHz. Inside this gap, a single narrow resonance appears at $f_{\mathrm{local}}=2.910$ GHz, close to the NV zero-field splitting $f_{\mathrm{ZFS}}=2.87$ GHz. The defect-supercell dispersion (Supplementary Fig.~S3) confirms that this state lies in a gap of the pristine crystal rather than on a folded bulk band, and the percell mode profile (Fig.~\ref{fig:field}a) confirms its confinement at the defect.

\subsection{Dynamic dipolar field at the NV site}

\begin{figure}[htbp]
\centering
\includegraphics[width=\textwidth]{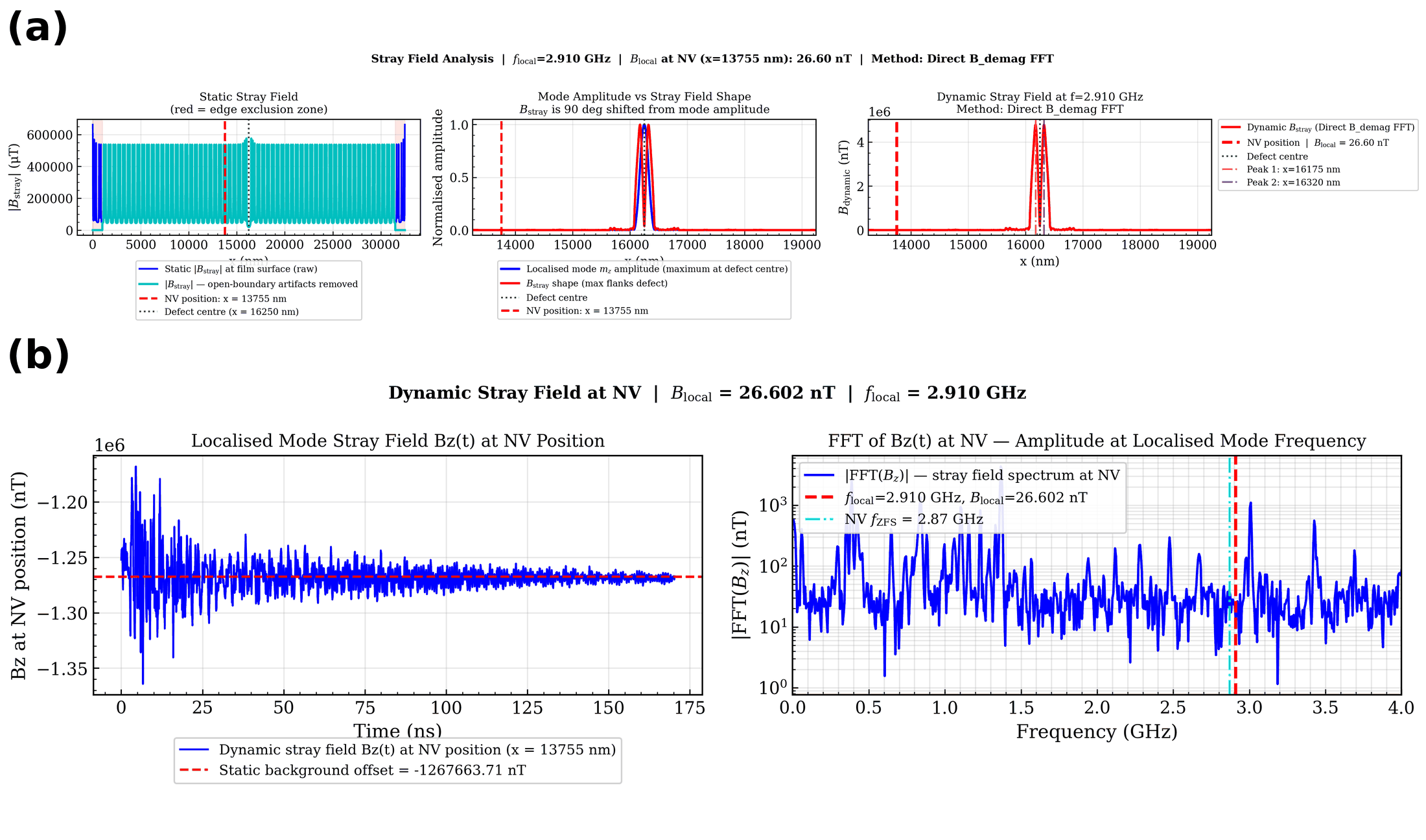}
\caption{\textbf{Localised mode and its dynamic field at the NV site.} (a) Left: static $|B_{\mathrm{stray}}|$ along the device, with edge-exclusion zones shaded. Middle: normalised localised-mode $m_{z}$ amplitude (blue) and dynamic stray-field envelope (red), showing the $\approx90^{\circ}$ spatial offset between the mode maximum at the defect centre and the field maxima on its flanks. Right: dynamic stray field at 2.910 GHz with its two flanking peaks. The red dashed line marks the NV position used here ($x=13{,}755$ nm). (b) Left: dynamic stray field $B_{z}(t)$ at the NV position, oscillating about the static background offset ($-1{,}267{,}664$ nT, red dashed line). Right: Fourier spectrum $|\mathrm{FFT}(B_{z})|$, giving $B_{\mathrm{local}}=26.60$ nT at $f_{\mathrm{local}}=2.910$ GHz (red dashed line), close to the NV zero-field splitting of 2.87 GHz (cyan dash-dotted line).}\label{fig:field}
\end{figure}

For NV detection, the relevant quantity is the dynamic dipolar field. The localised mode is produced at the NV positions, which the analysis takes directly from the MuMax3 $B_{\mathrm{demag}}$ output at each time step and Fourier transforms at each candidate position. Because the spatial gradient of the mode profile, rather than the mode amplitude, generates this field, $|B_{\mathrm{dyn}}(x)|$ peaks on the flanks of the defect, at $x\approx16{,}175$ and $16{,}320$ nm, rather than at its centre (Fig.~\ref{fig:field}a). The present analysis places the NV ensemble at $x=13{,}755$ nm, the position of the maximum static stray field within the search window (Supplementary Fig.~S2). At this position, $B_{z}(t)$ oscillates about a large static offset of $-1{,}267{,}664$ nT arising from the demagnetising field of the DC-biased crystal (Fig.~\ref{fig:field}b, left). After subtracting this offset, the Fourier spectrum gives $B_{\mathrm{local}}=26.60$ nT at $f_{\mathrm{local}}=2.910$ GHz (Fig.~\ref{fig:field}b, right). This value is the spectral amplitude of the response to the broadband pulse. Under continuous-wave (CW) excitation of the antenna at $f_{\mathrm{local}}$, the micromagnetic response is linear, and the ratio between the local dynamic field and the applied antenna field defines a magnonic field enhancement factor $\mathcal{E}=B_{\mathrm{local}}/B_{\mathrm{signal}}=0.3125$ ($B_{\mathrm{local}}=156$ $\mu$T for $B_{\mathrm{signal}}=0.5$ mT). Because the dynamic field peaks on the defect flanks, positions closer to them would give a substantially larger local field, a point revisited in the Discussion.

\subsection{Magnon-dressed NV spectra}

In the device, the antenna does not drive the NV centres directly; it excites the defect mode, whose transverse dipolar field drives the $m_{s}=0\leftrightarrow m_{s}=+1$ transition of each hyperfine sector. The drive therefore reaches the spins through the mode line. For a drive detuned by $\Delta_{d}=\omega_{d}-\omega_{c}$ from the mode, the Rabi frequency follows the Lorentzian amplitude response of the mode,
\begin{equation}
\Omega(\Delta_{d})=\frac{\Omega_{0}}{\sqrt{1+\left(2\Delta_{d}/\kappa_{m}\right)^{2}}},\qquad \frac{\Omega_{0}}{2\pi}=\gamma_{e}B_{\perp,\mathrm{local}},\label{eq:drive}
\end{equation}
where $\kappa_{m}$ is the full linewidth of the defect mode, $\gamma_{e}=28$ GHz/T and $B_{\perp,\mathrm{local}}$ is the component of the local dynamic field transverse to the NV axis when the drive is on the mode resonance. With $\mathcal{E}=0.3125$, the value $\Omega_{0}/2\pi=0.10$ MHz used below corresponds to $B_{\perp,\mathrm{local}}\approx3.6$ $\mu$T, that is, to an antenna field of about 11 $\mu$T; this keeps the spins in the moderately driven regime in which the three hyperfine lines stay resolved. The axes of the spectra are referred to the mode frequency, $\Delta_{s}=\omega_{s}-\omega_{c}$ and $\Delta_{d}=\omega_{d}-\omega_{c}$, as in ref.~\cite{kumar2026cavity}; in the rotating frame of Eq.~\eqref{eq:hamiltonian}, $\Delta_{c}=-\Delta_{d}$ and $\Delta_{j}=\Delta_{s}+m_{I}A_{\mathrm{hf}}-\Delta_{d}$.

\begin{figure}[htbp]
\centering
\includegraphics[width=\textwidth]{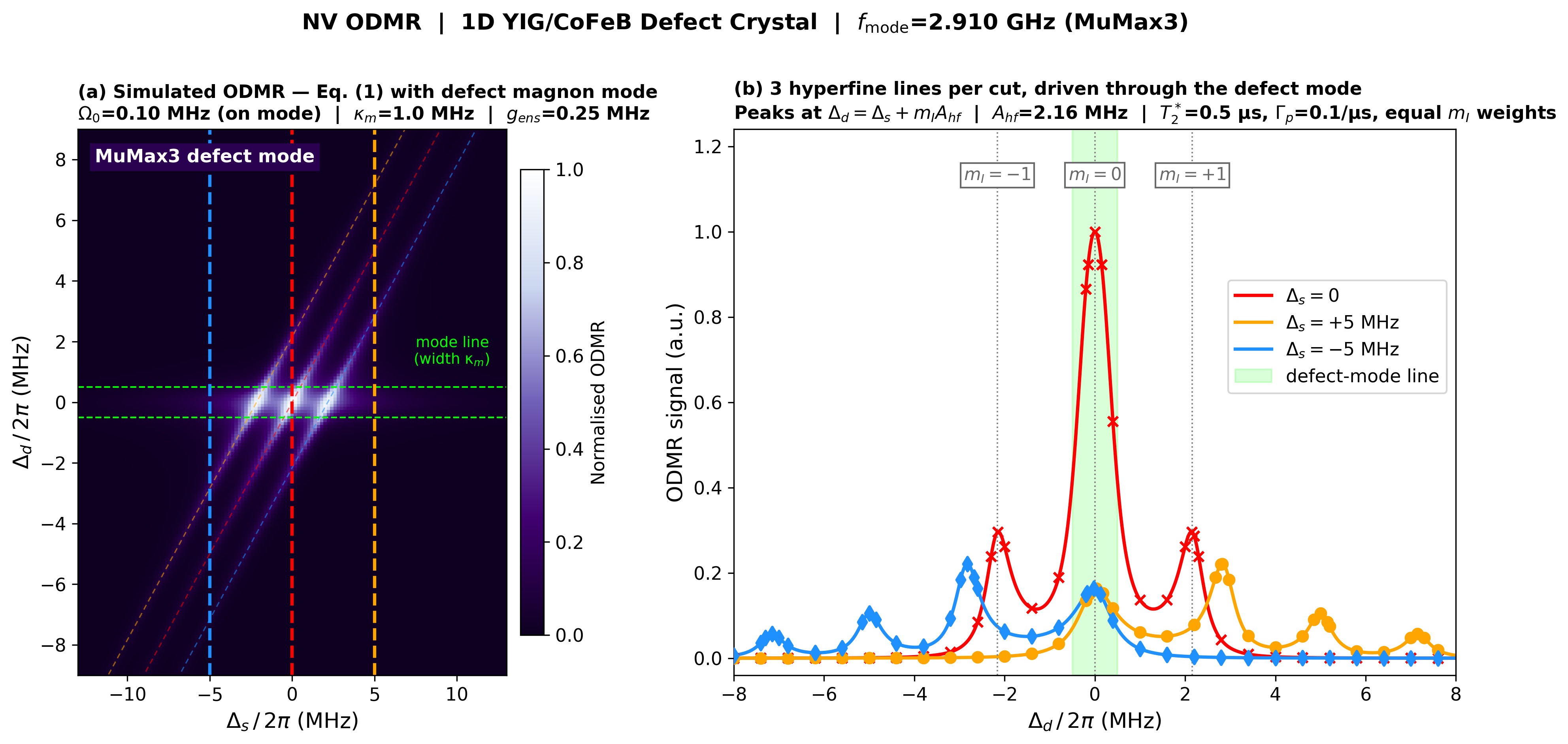}
\caption{\textbf{Simulated ODMR spectrum of the NV ensemble driven through the defect magnon mode.} (a) Normalised ODMR signal versus spin detuning $\Delta_{s}/2\pi$ and drive detuning $\Delta_{d}/2\pi$, both referred to the defect-mode frequency $f_{\mathrm{mode}}=2.910$ GHz, from the steady state of the Lindblad master equation [Eqs.~\eqref{eq:hamiltonian}--\eqref{eq:lindblad}] with the defect mode included and the drive filtered by the mode line [Eq.~\eqref{eq:drive}]. Dashed lines: hyperfine resonance conditions $\Delta_{d}=\Delta_{s}+m_{I}A_{\mathrm{hf}}$ ($m_{I}=+1$, 0, $-1$ in orange, red and blue). Green dashed lines: defect-mode line, $|\Delta_{d}|\leq\kappa_{m}/2$. Vertical dashed lines: positions of the cuts in (b). (b) Cuts at $\Delta_{s}/2\pi=0$ and $\pm5$ MHz; shaded band, defect-mode line. Parameters: $\Omega_{0}/2\pi=0.10$ MHz (on the mode resonance), $\kappa_{m}/2\pi=1.0$ MHz, $g_{\mathrm{ens}}/2\pi=0.25$ MHz, $A_{\mathrm{hf}}=2.16$ MHz, $T_{2}^{*}=0.5$ $\mu$s, $\Gamma_{p}=0.1$ $\mu$s$^{-1}$, $T_{1}=1$ ms and equal $m_{I}$ weights. We apply a single normalisation to the whole map.}\label{fig:odmr}
\end{figure}

Fig.~\ref{fig:odmr} shows the resulting ODMR spectrum, obtained from the steady state of Eqs.~\eqref{eq:hamiltonian}--\eqref{eq:lindblad} with the defect mode included ($\kappa_{m}/2\pi=1.0$ MHz, $g_{\mathrm{ens}}/2\pi=0.25$ MHz), an ensemble dephasing time $T_{2}^{*}=0.5$ $\mu$s, optical repolarisation into $m_{s}=0$ at $\Gamma_{p}=0.1$ $\mu$s$^{-1}$, $T_{1}=1$ ms and equal weights for the three $^{14}$N projections, which are equally populated at room temperature. The ODMR signal is the steady-state population transferred out of $m_{s}=0$, averaged over the three sectors and normalised once over the whole map. Three features stand out. First, the signal lies on three parallel ridges of unit slope, $\Delta_{d}=\Delta_{s}+m_{I}A_{\mathrm{hf}}$ with $A_{\mathrm{hf}}=2.16$ MHz (dashed lines), because the drive flips a spin only when its frequency matches that of the spin. Second, the ridges are strong only where they cross the defect-mode line ($|\Delta_{d}|\lesssim\kappa_{m}/2$, green band) and fade rapidly away from it, because the mode transmits the drive efficiently only near its own frequency (Eq.~\eqref{eq:drive}). Third, a weak horizontal band appears at $\Delta_{d}\approx0$ for all $\Delta_{s}$: there the drive is strongest, and even spins far from resonance are weakly excited off-resonantly. This band is not an additional resonance.

The line cuts in Fig.~\ref{fig:odmr}b make this quantitative. At $\Delta_{s}=0$, the central line, which lies inside the mode line, reaches unit height, while the two outer hyperfine lines at $\Delta_{d}/2\pi=\pm2.15$ MHz reach only 0.30, although all three sectors carry equal weight: they lie outside the 1 MHz-wide mode line and therefore receive a weaker drive. At $\Delta_{s}/2\pi=\pm5$ MHz, the three hyperfine lines appear at $\pm2.83$, $\pm5.00$ and $\pm7.15$ MHz with heights 0.22, 0.10 and 0.06, decreasing with distance from the mode, together with the off-resonant feature of height 0.16 near $\Delta_{d}=0$. A passive magnonic cavity therefore drives strongly only the hyperfine line that lies within its linewidth; a mode linewidth comparable to $2A_{\mathrm{hf}}$ would drive all three lines nearly equally. Only three distinct lines exist at any single bias field; the nine peaks of Fig.~\ref{fig:odmr}b are these three lines viewed at three bias fields. Resolving the Autler--Townes (Mollow) sidebands of ref.~\cite{kumar2026cavity} as additional ODMR features would require $\Omega$ to exceed the spin linewidth, a regime we do not pursue here.

For the microwave (homodyne) readout, a weak probe is reflected at the antenna frequency $\omega_{p}$, and the defect mode plays the role of the cavity mode $a$ from ref.~\cite{kumar2026cavity}. The ensemble coupling to the mode follows from the dynamic field per magnon,
\begin{equation}
g_{\mathrm{ens}}=\frac{\gamma_{e}B_{\mathrm{local}}}{\sqrt{n_{\mathrm{eff}}}},\label{eq:coupling}
\end{equation}
where $n_{\mathrm{eff}}$ is the number of magnons in the defect mode that produces the local field $B_{\mathrm{local}}$. A reliable value of $n_{\mathrm{eff}}$ requires the mode energy of the micromagnetic simulation; here $g_{\mathrm{ens}}/2\pi=0.25$ MHz is used as a representative value, the same as in Fig.~\ref{fig:odmr}, and its influence on the sensitivity is mapped in Fig.~\ref{fig:roadmap}. Eliminating the spin coherences from the quantum Langevin equations\cite{gardiner1985input} (Supplementary Note~3) gives the steady-state mode amplitude
\begin{equation}
\alpha=\frac{i\epsilon}{i\Delta_{c}+\gamma_{0}/2-g_{\mathrm{ens}}^{2}P\sum_{j}\left(i\Delta_{j}+\Gamma/2\right)^{-1}},\qquad \Delta_{j}=\Delta_{s}+m_{I}A_{\mathrm{hf}}+\Delta_{c},\label{eq:steady-state}
\end{equation}
where $\Delta_{c}=\omega_{c}-\omega_{p}$ is the detuning of the probe mode, which is swept, $\epsilon=\sqrt{\kappa_{\mathrm{ext}}}\,a^{\mathrm{mw}}$ is the probe amplitude coupled into the mode through the antenna ($\kappa_{\mathrm{ext}}$ is the antenna--mode coupling rate), $\gamma_{0}=\kappa_{m}$ is the total mode loss rate, $P\simeq-1$ is the spin polarisation maintained by optical pumping, and $\Gamma=2/T_{2}^{*}$ is the full spin linewidth ($\Gamma/2\pi=0.64$ MHz for $T_{2}^{*}=0.5$ $\mu$s). The factor $i$ fixes the quadrature phase so that $\mathrm{Re}\,\alpha>0$ is above the mode resonance. Because $\Delta_{j}$ contains $\Delta_{c}$, each spin maintains a fixed frequency while the probe is swept.

\begin{figure}[htbp]
\centering
\includegraphics[width=\textwidth]{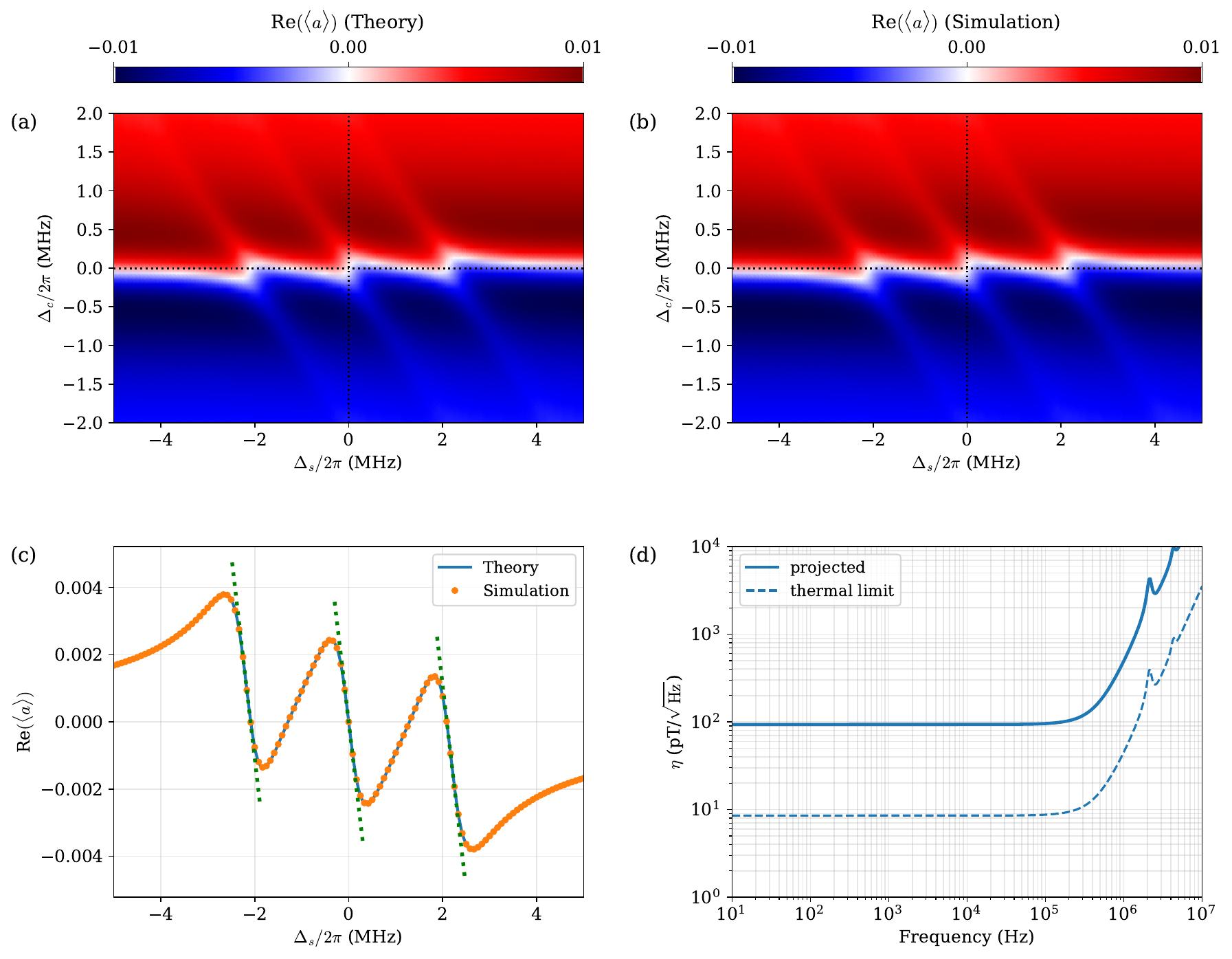}
\caption{\textbf{Homodyne response and projected sensitivity of the defect-mode readout.} (a) Steady-state reflected quadrature $\mathrm{Re}\langle a\rangle$ versus spin detuning $\Delta_{s}/2\pi$ and mode--probe detuning $\Delta_{c}/2\pi$ from the closed-form expression, Eq.~\eqref{eq:steady-state} (Theory). (b) The same quantity from the full Lindblad master equation (Simulation). (c) Cut at $\Delta_{c}=0$: theory (line) and simulation (markers); green dotted lines mark the steepest slopes used for the sensitivity. (d) Computed magnetic-noise spectral density $\eta(f)$ of the three edges combined, for an antenna critically coupled to the defect mode, a $-40$ dBm probe and room temperature, obtained from the frequency-dependent response of the reflected quadrature to an oscillating field (Methods). Dashed line: thermal (Johnson--Nyquist) limit; solid line: projected sensitivity including the technical-noise excess of the readout chain of ref.~\cite{kumar2026cavity}. The sensitivity is flat at 93 pT$/\sqrt{\mathrm{Hz}}$ (8.5 pT$/\sqrt{\mathrm{Hz}}$ thermal) up to about 100 kHz and rolls off above the 360 kHz bandwidth set by the spin and mode linewidths. Parameters: $\gamma_{0}/2\pi=\kappa_{m}/2\pi=1.0$ MHz, $g_{\mathrm{ens}}/2\pi=0.25$ MHz, $\Gamma/2\pi=0.64$ MHz, $\epsilon/2\pi=0.01$ MHz, $A_{\mathrm{hf}}=2.16$ MHz.}\label{fig:homodyne}
\end{figure}

Figs.~\ref{fig:homodyne}a,b show $\mathrm{Re}\langle a\rangle$ on the $(\Delta_{s},\Delta_{c})$ grid from the closed form of Eq.~\eqref{eq:steady-state} (Theory) and from the steady-state density matrix of the full master equation with mode, three spin sectors and a weak probe $\epsilon/2\pi=0.01$ MHz (Simulation), using the same parameters as Fig.~\ref{fig:odmr}. Both maps show the dispersive response of the defect mode, positive above and negative below $\Delta_{c}=0$ over a range set by $\gamma_{0}$, crossed by three tilted features where the hyperfine lines pull the mode frequency. The cut at $\Delta_{c}=0$ (Fig.~\ref{fig:homodyne}c) is a sawtooth with one steep edge per hyperfine line, at $\Delta_{s}/2\pi=-2.18$, 0 and $+2.18$ MHz, with nearly equal maximum slopes $|d\,\mathrm{Re}\langle a\rangle/d\Delta_{s}|\approx0.012$ per MHz (green dotted lines). These edges are the magnetometer's operating points: a field change shifts $\omega_{s}$, moves the sensor along an edge, and changes the reflected quadrature in proportion to the slope.

\subsection{Multispectral sensitivity}

The sensitivity for a single spectral feature follows the slope-detection framework of lock-in spin-resonance magnetometry~\cite{dreau2011,taylor2008,kumar2026cavity}. The homodyne quadrature $V_{Q}(\omega_{s})$ varies with the spin frequency $\omega_{s}$, which an external field $\delta B$ shifts through $\gamma_{e}$. At the steepest point of the $j$th edge, $\delta V_{Q}^{(j)}=S_{j}\gamma_{e}\delta B$ with
\begin{equation}
  S_j = \left.\left|\frac{dV_Q}{d\omega_s}\right|\right._{\!\max}^{(j)},
\end{equation}
and with r.m.s. quadrature noise $\sigma_{j}$ and integration time $\tau$, the single-feature sensitivity is
\begin{equation}
  \eta_{j} = \frac{\sqrt{3}\,\sigma_{j}\sqrt{\tau}}{\gamma_{e}S_{j}}.
  \label{eq:eta-j}
\end{equation}
The prefactor $\sqrt{3}$ is retained from ref.~\cite{kumar2026cavity} for direct comparison; it describes the projection $\cos54.7^{\circ}=1/\sqrt{3}$ of a bias field along $\langle100\rangle$ onto each NV axis, and for randomly oriented nanodiamonds the orientation average changes it by a factor of order unity. Features that are resolved at the same operating point give independent estimates of $\delta B$ and combine in quadrature,
\begin{equation}
  \eta_{\mathrm{multi}} =
  \left(\sum_{j=1}^{M}\eta_{j}^{-2}\right)^{-1/2},
  \label{eq:eta-multi}
\end{equation}
where $M$ counts the edges present in one bias field. For the device in defect-mode, $M=3$ (Fig.~\ref{fig:homodyne}c).

Without a fabricated device, we cannot measure $\sigma_{j}$, so we project the sensitivity in two steps (Methods). First, the slope of the reflected quadrature, calculated from Eq.~\eqref{eq:steady-state} for a critically coupled antenna ($\kappa_{\mathrm{ext}}=\kappa_{m}/2$) and a probe of $-40$ dBm, is divided by the thermal noise (Johnson--Nyquist) of a room-temperature readout\cite{clerk2010introduction}. This gives a thermal limit of 14.7 pT$/\sqrt{\mathrm{Hz}}$ for each edge and 8.5 pT$/\sqrt{\mathrm{Hz}}$ for the three combined. Second, the same calculation for the parameters of ref.~\cite{kumar2026cavity} gives 1.1 pT$/\sqrt{\mathrm{Hz}}$, against the measured 12 pT$/\sqrt{\mathrm{Hz}}$; a practical homodyne chain adds a technical-noise excess of about 11. Applying the same excess gives projected sensitivities of 161 pT$/\sqrt{\mathrm{Hz}}$ for the best single edge and $\eta_{\mathrm{multi}}=93$ pT$/\sqrt{\mathrm{Hz}}$ for the three edges, a multiplexing gain of 1.72, close to the $\sqrt{3}$ expected for three edges of nearly equal slope. Fig.~\ref{fig:homodyne}d shows how this sensitivity depends on the frequency of the measured field. A field oscillating at frequency $f$ modulates the spin frequencies, and the coupled mode--spin system follows the modulation only up to a rate set by its linewidths. The computed $\eta(f)$ is therefore flat at the projected floor from the lowest frequencies up to about 100 kHz, which covers the band relevant to magnetic-anomaly detection and geomagnetic navigation, and rises above a $-3$ dB bandwidth of about 360 kHz. The small features near 2 MHz appear where the modulation sidebands reach the neighbouring hyperfine lines. A real readout would add $1/f$ noise and pickup at low frequencies, which the model does not include. The antenna coupling $\kappa_{\mathrm{ext}}$ enters directly and is analysed in Fig.~\ref{fig:roadmap}b: the best readout is obtained when the antenna is matched to the total internal loss of the mode at the operating point, which includes the absorption by the spins, and the sensitivity degrades rapidly for weak antenna coupling\cite{pozar2012microwave}. The coupling of the antenna to the nanoscale defect mode is, therefore, a key quantity for an experimental device.

\subsection{Theory--simulation validation}

Because the sensitivity rests on slopes extracted from the numerical master-equation solution, we checked each route against an independent analytic treatment.

\subsubsection{ODMR route}

\begin{figure}[htbp]
\centering
\includegraphics[width=\textwidth]{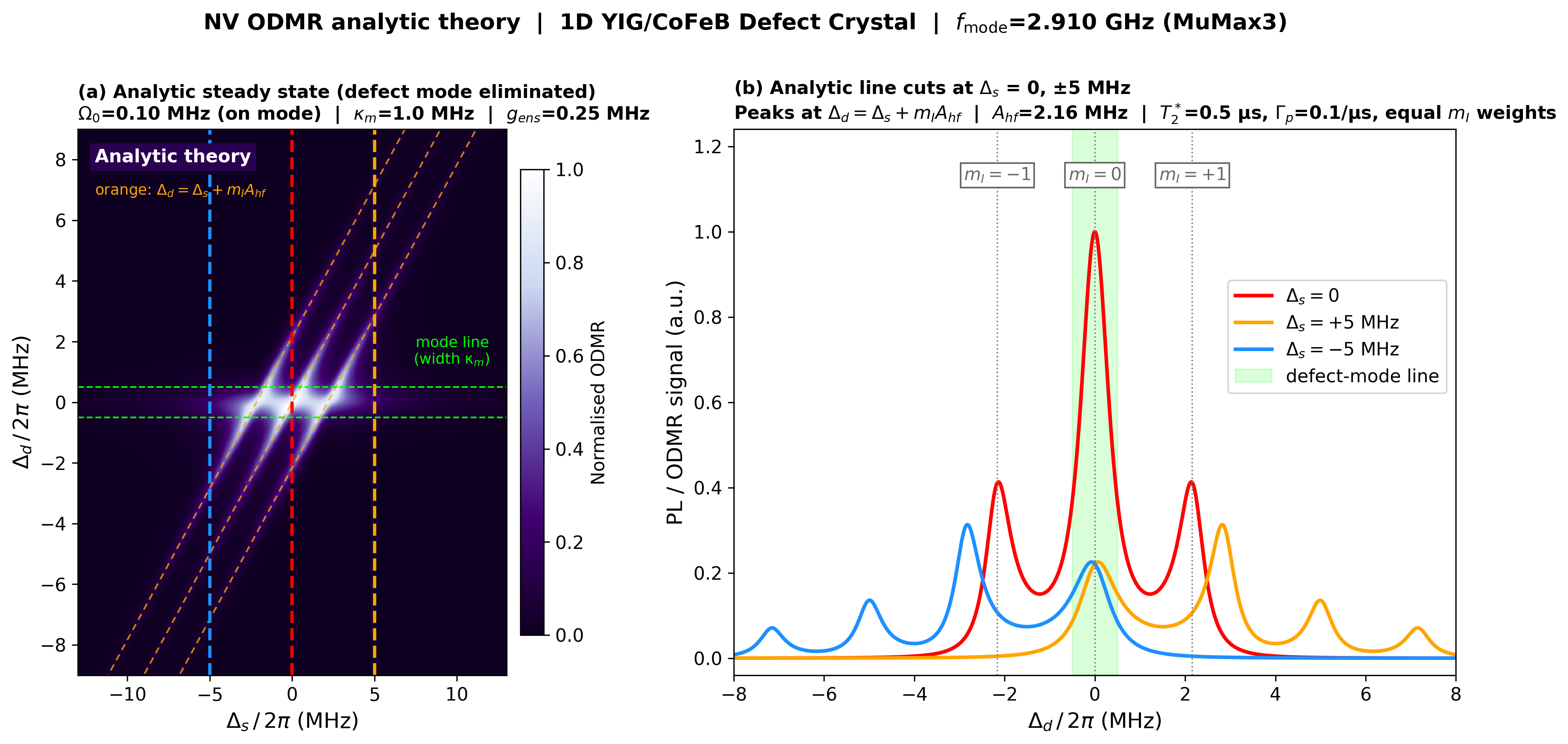}
\caption{\textbf{Analytic ODMR theory of the defect-mode-driven NV ensemble.} (a) Normalised ODMR signal versus $\Delta_{s}/2\pi$ and $\Delta_{d}/2\pi$ from the analytic steady state (Bloch equations for each hyperfine sector with the drive of Eq.~\eqref{eq:drive}, plus the Purcell decay and frequency pull of the adiabatically eliminated defect mode). Orange dashed lines: $\Delta_{d}=\Delta_{s}+m_{I}A_{\mathrm{hf}}$; green dashed lines: defect-mode line; vertical dashed lines: cut positions. (b) Analytic line cuts at $\Delta_{s}/2\pi=0$ and $\pm5$ MHz. Parameters as in Fig.~\ref{fig:odmr}.}\label{fig:theory-overlay}
\end{figure}

Fig.~\ref{fig:theory-overlay} shows the ODMR spectrum from an approximate analytic steady state (Supplementary Note~3). Each hyperfine sector is treated as a driven two-level system with the drive of Eq.~\eqref{eq:drive}; the defect mode is eliminated adiabatically, which adds a Purcell decay $\Gamma_{\mathrm{P}}=g_{\mathrm{ens}}^{2}\gamma_{0}/(\Delta_{sj}^{2}+\gamma_{0}^{2}/4)$ and a mode-induced frequency pull $\delta_{\mathrm{L}}=g_{\mathrm{ens}}^{2}\Delta_{sj}/(\Delta_{sj}^{2}+\gamma_{0}^{2}/4)$ to each sector, with $\Delta_{sj}=\Delta_{s}+m_{I}A_{\mathrm{hf}}$ the spin--mode detuning. The analytical map reproduces the three ridges, their confinement to the mode line and the off-resonant band at $\Delta_{d}\approx0$ (Fig.~\ref{fig:theory-overlay}a). In the cuts (Fig.~\ref{fig:theory-overlay}b), the peak positions agree with the simulation of Fig.~\ref{fig:odmr}b to within 0.012 MHz r.m.s., and the analytic ridges follow the lines $\Delta_{d}=\Delta_{s}+m_{I}A_{\mathrm{hf}}$ (orange). The analytic heights of the outer lines are larger than the simulated ones (0.41 instead of 0.30 at $\Delta_{s}=0$, and 0.31 instead of 0.22 at $\Delta_{s}/2\pi=\pm5$ MHz), and the agreement on the whole map is $R^{2}=0.91$. The deviation is highest near the mode line, where the adiabatic elimination is less accurate because $g_{\mathrm{ens}}$ is not small compared to $\gamma_{0}/2$ and the three sectors share one mode. It decreases as $g_{\mathrm{ens}}$ decreases, and the closed form reproduces the master equation exactly when the mode coupling is off. The analytic theory therefore captures the positions and the overall pattern, while quantitative peak heights require the full master equation.

\subsubsection{Homodyne route}

For the homodyne response, the closed form of Eq.~\eqref{eq:steady-state} and the Lindblad solution agree point by point over the whole map (Fig.~\ref{fig:homodyne}a,b), with $R^{2}>0.9999$ and a maximum deviation below $10^{-4}$ of the peak signal; the cut at $\Delta_{c}=0$ (Fig.~\ref{fig:homodyne}c) agrees to an r.m.s. error of $2\times10^{-7}$. This agreement is expected in the weak-probe regime, where the spins stay polarised and Eq.~\eqref{eq:steady-state} is exact for the model, and it confirms that the slopes $S_{j}$ entering Eq.~\eqref{eq:eta-j} are free of numerical artefacts. Table~\ref{tab:validation} summarises the checks.

\begin{table}[htbp]
\caption{Summary of the theory--simulation validation and of the projected sensitivity. Positions and heights refer to the cuts of Figs.~\ref{fig:odmr}b and \ref{fig:theory-overlay}b; the homodyne values refer to Fig.~\ref{fig:homodyne}.}\label{tab:validation}
\begin{tabular}{@{}lccc@{}}
\toprule
Quantity & Simulation & Analytic & Comment\\
\midrule
ODMR peak positions (r.m.s. difference) & -- & -- & 0.012 MHz\\
Outer hyperfine line height, $\Delta_{s}=0$ & 0.30 & 0.41 & drive filtered by mode line\\
ODMR map agreement, $R^{2}$ & -- & -- & 0.91\\
Homodyne map agreement, $R^{2}$ & -- & -- & $>0.9999$\\
Homodyne edge positions, $\Delta_{s}/2\pi$ & $-2.18$, 0, $+2.18$ MHz & identical & one per hyperfine line\\
Thermal limit, three edges & \multicolumn{2}{c}{8.5 pT$/\sqrt{\mathrm{Hz}}$} & critical coupling, $-40$ dBm, 300 K\\
Projected best single edge, $\eta_{\mathrm{single}}$ & \multicolumn{2}{c}{161 pT$/\sqrt{\mathrm{Hz}}$} & technical excess $\times$11\\
Projected three-edge sensitivity, $\eta_{\mathrm{multi}}$ & \multicolumn{2}{c}{93 pT$/\sqrt{\mathrm{Hz}}$} & Eq.~\eqref{eq:eta-multi}, $M=3$\\
Multiplexing gain, $\eta_{\mathrm{single}}/\eta_{\mathrm{multi}}$ & \multicolumn{2}{c}{1.72} & $\sqrt{3}=1.73$ for equal edges\\
\botrule
\end{tabular}
\end{table}

\subsection{Design levers and projected sensitivity}

\begin{figure}[htbp]
\centering
\includegraphics[width=\textwidth]{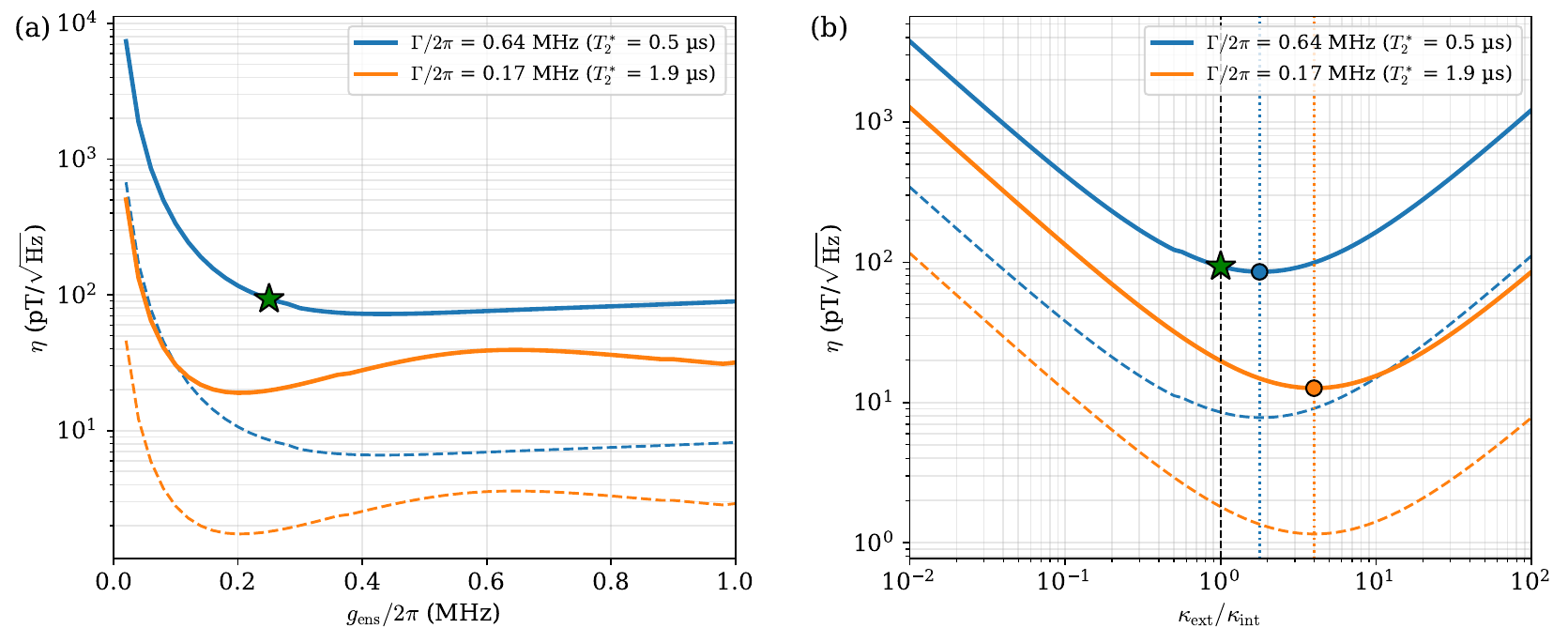}
\caption{\textbf{Design levers of the projected sensitivity.} Three-edge sensitivity $\eta_{\mathrm{multi}}$ for a $-40$ dBm probe at room temperature and spin linewidths $\Gamma/2\pi=0.64$ MHz (blue) and 0.17 MHz (orange). Solid lines: projected sensitivity, including the technical-noise excess of ref.~\cite{kumar2026cavity}; dashed lines: thermal (Johnson--Nyquist) limit. (a) Versus NV--magnon coupling $g_{\mathrm{ens}}/2\pi$ for an antenna critically coupled to the defect mode ($\kappa_{m}/2\pi=1.0$ MHz). (b) Versus antenna--mode coupling $\kappa_{\mathrm{ext}}/\kappa_{\mathrm{int}}$ for $\kappa_{\mathrm{int}}/2\pi=0.5$ MHz and $g_{\mathrm{ens}}/2\pi=0.25$ MHz; black dashed line, critical coupling; circles, computed optima; dotted lines, $\kappa_{\mathrm{ext}}=\kappa_{\mathrm{int}}+4g_{\mathrm{ens}}^{2}/\Gamma$. The green stars mark the parameters of Figs.~\ref{fig:odmr}--\ref{fig:homodyne} (93 pT$/\sqrt{\mathrm{Hz}}$).}\label{fig:roadmap}
\end{figure}

Fig.~\ref{fig:roadmap}a maps the projected three-edge sensitivity as a function of $g_{\mathrm{ens}}$ for two spin linewidths, with the antenna critically coupled ($\kappa_{\mathrm{ext}}=\kappa_{\mathrm{int}}=\kappa_{m}/2$). For $\Gamma/2\pi=0.64$ MHz ($T_{2}^{*}=0.5$ $\mu$s), the sensitivity improves steeply to $g_{\mathrm{ens}}/2\pi\approx0.3$ MHz and reaches an optimum of 72 pT$/\sqrt{\mathrm{Hz}}$ near 0.44 MHz; beyond the optimum, the coupling begins to split the features and the slopes no longer grow. Narrowing the spin linewidth to 0.17 MHz ($T_{2}^{*}=1.9$ $\mu$s), the value used in~\cite{kumar2026cavity}, improves the optimum to 19 pT$/\sqrt{\mathrm{Hz}}$ near $g_{\mathrm{ens}}/2\pi=0.20$ MHz. The dashed curves give the thermal limit of each case, 6.6 and 1.7 pT$/\sqrt{\mathrm{Hz}}$ at the respective optimals, which a readout chain without technical excess would approach.

Fig.~\ref{fig:roadmap}b varies the antenna--mode coupling $\kappa_{\mathrm{ext}}$ at fixed internal mode loss ($\kappa_{\mathrm{int}}/2\pi=0.5$ MHz) and $g_{\mathrm{ens}}/2\pi=0.25$ MHz. The sensitivity has a clear optimum, but not at the textbook critical coupling $\kappa_{\mathrm{ext}}=\kappa_{\mathrm{int}}$ of an empty resonator. At the operating point, the spins absorb energy from the mode at a rate of about $4g_{\mathrm{ens}}^{2}/\Gamma$, which adds to its internal loss, and the antenna is best matched to this total loss, $\kappa_{\mathrm{ext}}^{\mathrm{opt}}\approx\kappa_{\mathrm{int}}+4g_{\mathrm{ens}}^{2}/\Gamma$ (dotted lines). This gives $\kappa_{\mathrm{ext}}^{\mathrm{opt}}=1.8\,\kappa_{\mathrm{int}}$ and 85 pT$/\sqrt{\mathrm{Hz}}$ for $\Gamma/2\pi=0.64$ MHz, 9\% better than critical coupling, and $\kappa_{\mathrm{ext}}^{\mathrm{opt}}=4.0\,\kappa_{\mathrm{int}}$ and 13 pT$/\sqrt{\mathrm{Hz}}$ for $\Gamma/2\pi=0.17$ MHz, where the spin absorption is stronger. Away from the optimum the sensitivity degrades on both sides; an antenna coupled twenty times more weakly than $\kappa_{\mathrm{int}}$ loses a factor of 8--13, more than the empty-resonator estimate, because the spin absorption increases the mismatch. A slightly over-coupled antenna is therefore preferable to an under-coupled one.

%% =====================================================================
\section{Discussion}

This work shows that the cavity-coupled sensing scheme behind cavity-enhanced NV magnetometry\cite{kumar2026cavity} carries over to a passive magnon mode, within the same Tavis--Cummings/Lindblad model. A defect-engineered 1D YIG/CoFeB magnonic crystal supports a localised spin-wave mode inside its band gap at 2.910 GHz. When the microwave reaches the NV ensemble through this mode, the ODMR spectrum is hyperfine-resolved but strongly weighted toward the line within the mode linewidth, and the homodyne readout of the mode provides one steep dispersive edge per hyperfine line. Parameter-free checks against a plane-wave band-structure calculation and against closed-form steady states make each step traceable to the established theory.

Quantitatively, the projected three-edge sensitivity of 93 pT$/\sqrt{\mathrm{Hz}}$ is approximately eight times lower than the 12 pT$/\sqrt{\mathrm{Hz}}$ measured with the driven dielectric cavity, and the multiplexing gain of 1.72 is limited by the three edges available in a single bias field, compared with the eight or nine features of the strongly driven cavity. The analysis identifies the levers that close this gap: (i) the antenna--mode coupling, which should match the total loss of the mode including the spin absorption, so that a slightly over-coupled antenna is optimal (Fig.~\ref{fig:roadmap}b); (ii) the NV--magnon coupling $g_{\mathrm{ens}}$, set by the NV-to-film stand-off and by placing the ensemble on the defect flanks, where the dynamic field is largest (Fig.~\ref{fig:field}a), with an optimum near $g_{\mathrm{ens}}/2\pi\approx0.2$--0.4 MHz (Fig.~\ref{fig:roadmap}a); (iii) the spin linewidth, where higher-quality diamond improves the projection to about 19 pT$/\sqrt{\mathrm{Hz}}$; and (iv) a defect-mode linewidth comparable to $2A_{\mathrm{hf}}$, so that all three hyperfine lines are driven equally, together with lower Gilbert damping\cite{gilbert2004phenomenological} of the localised mode. Unlike a driven cavity, fabrication fixes these levers, with no continuous high-power microwave drive, which favours size, weight, power, and thermal budget in chip-scale deployment.

To the best of the authors' knowledge, this is the first physically grounded simulation showing that cavity-coupled readout in cavity-driven NV magnetometry survives replacing an actively driven photon mode with a passive, defect-localised magnon mode. The result opens a design space, termed here \textbf{magnon-mediated multispectral sensing}, that complements and can be tested against its cavity-driven counterpart. Natural next steps include multi-defect crystals, analogous to coupled-cavity arrays,\cite{kumar2022enhanced,kumar2026spin} for independent multiplexed readout, self-consistent micromagnetic--QuTiP co-simulation including magnon--NV back-action, and experimental verification of the predicted localised mode in a fabricated crystal integrated with a diamond NV layer. For MAD and passive geomagnetic navigation,\cite{yu2023underwater,yu2025experimental} a cavity-free, low-power sensor head is attractive even though MAD detection range grows only as the cube root of any sensitivity improvement\cite{coy2025assessing}.

%% =====================================================================
\section{Methods}

\subsection{Micromagnetic simulations}

All micromagnetic simulations used the GPU-accelerated finite-difference code MuMax3\cite{vansteenkiste2014design}, which integrates the Landau--Lifshitz--Gilbert (LLG) equation
\begin{equation}
\frac{\partial\mathbf{m}}{\partial t}=-\frac{\gamma}{1+\alpha^{2}}\left[\mathbf{m}\times\mathbf{B}_{\mathrm{eff}}+\alpha\,\mathbf{m}\times\left(\mathbf{m}\times\mathbf{B}_{\mathrm{eff}}\right)\right],\qquad
\mathbf{B}_{\mathrm{eff}}=\mathbf{B}_{\mathrm{ext}}+\mathbf{B}_{\mathrm{exch}}+\mathbf{B}_{\mathrm{demag}},\label{eq:llg}
\end{equation}
for the reduced magnetisation $\mathbf{m}=\mathbf{M}/M_{s}$ in each cell, where $\gamma$ is the gyromagnetic ratio, $\alpha$ the Gilbert damping constant and $\mathbf{B}_{\mathrm{eff}}$ the effective field, which contains the applied field (Zeeman), the exchange field $\mathbf{B}_{\mathrm{exch}}=(2A_{\mathrm{ex}}/M_{s})\nabla^{2}\mathbf{m}$ and the demagnetising field (dipolar) computed by fast Fourier convolution over the whole sample. Magnetocrystalline anisotropy is neglected for both materials. The one-dimensional magnonic crystal consists of alternating YIG and CoFeB segments of widths $ s=300$ nm and $ d=200$ nm (lattice constant $a=s+d$) in a continuous film of thickness 25 nm on a non-magnetic GGG substrate, with a total length of about 32 $\mu$m along the periodicity direction $x$. A single point defect, formed by lengthening one YIG segment, is centred at $x=16{,}250$ nm, so that about 32 unit cells of the pristine crystal lie on either side of it and confine the defect mode. The sample is discretised into cells of $\Delta x\times\Delta y\times\Delta z=5 \times 5\times 5\times \mathrm{nm}^{3}$, smaller than the exchange lengths $\lambda_{\mathrm{ex}}=\sqrt{2A_{\mathrm{ex}}/\mu_{0}M_{s}^{2}}$ of both materials, so that the exchange coupling across the YIG/CoFeB interfaces and the short-wavelength modes near the band edges are resolved. Each cell is assigned the saturation magnetisation $M_{s}$, exchange stiffness $A_{\mathrm{ex}}$ and damping $\alpha$ of its material (Supplementary Table~S1), and MuMax3 couples neighbouring regions through the harmonic mean of their exchange stiffnesses. There are no periodic boundary conditions applied, so the domain has open boundaries.

The stray field of the on-chip permanent magnets is represented by a uniform bias $\mathbf{B}_{\mathrm{ext}}=101.4$ mT along $x$, parallel to the direction of periodicity and to the spin-wave vector, which places the crystal in backward volume geometry. Starting from a uniform magnetisation along $x$, the relax() routine minimises the total micromagnetic energy, the sum of the Zeeman, exchange, and demagnetising energies, with a torque-based steepest-descent scheme, and stops when the maximum torque falls below the solver's tolerance; no time integration takes place at this stage. The relaxed state is nearly saturated along $x$, with small canting at the YIG/CoFeB interfaces and at the film ends, where magnetic surface charges produce a strongly non-uniform internal field. The static demagnetising field of this state, read from the $B_{\mathrm{demag}}$ output, gives the stray-field maps of Supplementary Figs.~S1 and S2: a periodic dipolar pattern with a magnitude of several hundred millitesla at the film surface and a local dip at the defect centre. The open boundaries produce demagnetising artefacts within about 1 $\mu$m of each enwe exclude these zones excluded from all field and mode analysis.

The spin-wave dynamics were then obtained from the relaxed state in two complementary ways. For the spectra and dispersions, a spatially uniform sinc-pulse field of 10 mT amplitude and 5 GHz cut-off, applied perpendicular to the bias, excites all modes below the cut-off with comparable amplitude. The ring-down is recorded for about 350 ns (Fig.~\ref{fig:time-traces}) at a sampling interval of 10 ps, below the 100 ps Nyquist limit of the 5 GHz band, with the physical Gilbert damping constants
$\alpha_{\mathrm{YIG}} = 2\times10^{-4}$ and $\alpha_{\mathrm{CoFeB}} = 5\times10^{-3}$, which set the spectral linewidths without shifting the mode frequencies; the magnetisation is stored both as a volume average and cell by cell along $x$. The Fourier transform of $\langle m_{z}(t)\rangle$ gives the spectral power $S(f)$ of Fig.~\ref{fig:bandgap}b, and a two-dimensional space--time Fourier transform of the cell-resolved record $m_{z}(x,t)$ gives the dispersion of Fig.~\ref{fig:bandgap}a; for the crystal with the defect, the whole domain is treated as one supercell (Supplementary Note~2). The dynamic dipolar field at candidate NV positions, on the film surface, is read directly from the $B_{\mathrm{demag}}$ output at every time step; its large static offset is subtracted and its Fourier amplitude at the defect-mode frequency $f_{\mathrm{local}}=2.910$ GHz is evaluated over a search window of $\pm5$ unit cells around the defect, which fixed the NV position at $x=13{,}755$ nm. To calibrate the drive strength, a continuous-wave field of amplitude $B_{\mathrm{signal}}=0.5$ mT at $f_{\mathrm{local}}$ is applied until the response reaches a steady state; because the response is linear at this amplitude, the resulting local dynamic field at the NV position, $B_{\mathrm{local}}=156$ $\mu$T, defines the magnonic field enhancement factor $\mathcal{E}=B_{\mathrm{local}}/B_{\mathrm{signal}}=0.3125$, which relates the Rabi frequency of Eq.~\eqref{eq:drive} to the antenna field.

\subsection{Band-structure calculations}

The plane-wave method (PWM) provides an independent calculation of the backward-volume band structure of the pristine crystal, following the formulation developed for striped Co/permalloy magnonic waveguides\cite{kumar2013spin} and the plane-wave
theory of magnonic crystals\cite{krawczyk2008plane, puszkarski2003magnonic,vasseur1996magnon}. The linearised Landau--Lifshitz equation is written for small deviations $\mathbf{m}=(0,m_{y},m_{z})$ from static magnetisation along the bias $\mathbf{M}\approx M_{s}(x)\hat{\mathbf{x}}$, with harmonic time dependence $e^{i\omega t}$. The periodic material parameters, $M_{s}(x)$ and $\lambda_{\mathrm{ex}}^{2}(x)=2A_{\mathrm{ex}}/\mu_{0}M_{s}^{2}$, are expanded in Fourier series over the reciprocal lattice vectors $G_{n}=2\pi n/a$, and by Bloch's theorem the dynamic magnetisation is written as $m_{y,z}(x)=\sum_{n}m_{y,z}(G_{n})\,e^{i(k+G_{n})x}$, with the Bloch wave vector $k$ in the first Brillouin zone. The magnetostatic potential of the periodic magnetisation in the 25 nm film is obtained from the magnetostatic potential of the periodic magnetisation in the 25 nm film, averaged over the thickness of the film; effective dipolar pinning at the lateral edges quantises the transverse wave vector.\cite{guslienko2002effective} Substituting these expansions turns the equation of motion into the matrix eigenvalue problem $\hat{M}\mathbf{m}_{k}=i\omega\mathbf{m}_{k}$ of size $2N\times2N$ for reciprocal lattice vectors $N$, which is diagonalised on a grid of $k$ Positive eigenvalues give the bands $f_{n}(k)=\omega_{n}(k)/2\pi$, and the eigenvectors give the mode profiles. The calculation truncates the expansion at $N = 21$
harmonics ($2N+1 = 43$ reciprocal lattice vectors, an $86\times86$ eigenvalue problem); increasing $N$ further changes the bands below 4~GHz by less than
1\%. \\
\noindent All material parameters are taken directly from the micromagnetic model, without fitting to its output. Supplementary Note~2 gives the full formalism, the
sub-matrices, and a cheque against the unifobackwards-volumed-volume dispWe obtained the micromagnetic dispersion obtained without assuming any band model, from the cell-resolved ring-down $m_{z}(x,t)$ after the broadband pulse. A two-dimensional Fourier transform, first from position to wave vector and then from time to frequency, gives the dynamical structure factor $S(k,f)=|\mathcal{F}_{x,t}[m_{z}(x,t)]|^{2}$, whose intensity ridges trace the spin-wave bands; a Hann window in time and the exclusion of the edge zones suppress leakage and boundary artefacts, and the result is plotted on a logarithmic scale against the normalised wave vector $kd$ (Fig.~\ref{fig:bandgap}a). For the pristine crystal, the PWM bands are overlaid on this map without adjustment. For the crystal containing the defect, the single-cell PWM no longer applies because the discrete translational symmetry is broken. The whole simulated domain is then treated as one supercell of length $L$, which folds the bulk bands into closely spaced, nearly flat replica branches separated by about $2\pi/L$ (Supplementary Fig.~S3)\cite{venkat2014micromagnetic,kumar2022enhanced}. A mode is classified as defect-localised only if two independent criteria hold: its frequency lies inside a gap of the pristine band structure rather than on a folded bulk band, and its spatial profile, taken from the per-cell Fourier amplitude at that frequency, decays away from the defect (Fig.~\ref{fig:field}a). The mode at 2.910 GHz meets both criteria.

\subsection{Master-equation simulations}

QuTiP\cite{johansson2012qutip} solves Eqs.~\eqref{eq:hamiltonian} and \eqref{eq:lindblad} for one bosonic mode (truncated at two Fock states for the ODMR route and three for the homodyne route) and three hyperfine sectors with $A_{\mathrm{hf}}=2.16$ MHz. Collapse operators describe the mode loss $\sqrt{\gamma_{0}}\,a$, spin relaxation and optical repolarisation $\sqrt{T_{1}^{-1}+\Gamma_{p}}\,\sigma_{j}^{-}$ and pure dephasing $\sqrt{\gamma_{\phi}/2}\,\sigma_{j}^{z}$ with $\gamma_{\phi}=1/T_{2}^{*}-1/(2T_{1})$; all rates are in $\mu$s$^{-1}$ and all frequencies in rad\,$\mu$s$^{-1}$. For the ODMR route, the drive enters through Eq.~\eqref{eq:drive}, the observable is the mean population of $|m_{s}=+1\rangle$ over the three sectors with equal weights, and the whole map is normalised once. For the homodyne route, a weak probe $\epsilon(a+a^{\dagger})$ drives the mode, the spins are kept polarised by pumping, and the observable is $\mathrm{Re}\,\mathrm{Tr}(a\rho_{\mathrm{ss}})$. Supplementary Note~4 lists all parameters.

\subsection{Sensitivity projection}

The reflection coefficient of the antenna at $\Delta_{c}=0$ is $r=1-\kappa_{\mathrm{ext}}\alpha/\epsilon$, with $\alpha$ from Eq.~\eqref{eq:steady-state} and $\kappa_{\mathrm{ext}}=\kappa_{m}/2$ for critical coupling. For an input power $P_{\mathrm{in}}$, a change in the field moves the reflected quadrature by $\sqrt{P_{\mathrm{in}}}\,|dX/d\omega_{s}|\,\gamma_{e}\delta B$, where $X$ is the dispersive quadrature of $r$. The thermal (Johnson--Nyquist) noise of one quadrature has the one-sided spectral density $S_{b}=(k_{B}T+hf/2)/2$\cite{clerk2010introduction}, which gives the thermal limit $\eta_{j}^{\mathrm{th}}=\sqrt{3}\sqrt{S_{b}}/(\gamma_{e}\sqrt{P_{\mathrm{in}}}\,|dX/d\omega_{s}|_{\max,j})$ for each edge, with $P_{\mathrm{in}}=-40$ dBm and $T=300$ K. The technical-noise excess of a practical readout chain is taken from ref.~\cite{kumar2026cavity} as the ratio of its measured sensitivity (12 pT$/\sqrt{\mathrm{Hz}}$) to the thermal limit computed in the same way for its parameters ($\kappa/2\pi=0.25$ MHz assumed, $g_{\mathrm{ens}}/2\pi=0.26$ MHz, $\Gamma/2\pi=167$ kHz, nine features), which gives 10.9. For the spectrum of Fig.~\ref{fig:homodyne}d, the mean-field Langevin equations are linearised about the steady state at each edge, with the spin detunings modulated as $\Delta_{j}\to\Delta_{j}+\gamma_{e}\delta B\cos(2\pi ft)$. Solving the resulting linear system at each modulation frequency gives the quadrature response $Y_{j}(f)$, which reduces to the static slope as $f\to0$, and $\eta_{j}(f)=\eta_{j}(0)\,|Y_{j}(0)|/|Y_{j}(f)|$; the three edges are combined with Eq.~\eqref{eq:eta-multi}. The output noise is the thermal noise of the room-temperature line, white over the band considered (Supplementary Note~5). The projection assumes that the spins respond linearly at the probe power; this requires a sufficiently large ensemble and should be verified for a specific device.

%% =====================================================================
\backmatter
\section*{Backmatter}
\addcontentsline{toc}{section}{Backmatter}

\section*{Supplementary information}
 Supplementary information contains Supplementary Notes 1--5, Supplementary Figs.~S1--S3 and Supplementary Tables~S1--S2.

\subsection*{Acknowledgements}
The author acknowledges the Centre for Computational Modelling and Simulation (CCMS), National Institute of Technology Calicut, India, for access to the Madhava
High-Performance Computing (HPC) cluster, and the Department of Electronics and Communication Engineering, National Institute of Technology Calicut, for the
departmental workstation on which part of the simulations were carried out.

\phantomsection
\pdfbookmark[1]{Declarations}{declarations}
\section*{\texorpdfstring{Declarations}{Declarations}}
\addcontentsline{toc}{section}{Declarations}

\bmhead{Declarations}

\pdfbookmark[1]{Funding}{funding}
%\section*{Funding}
This research did not receive a specific grant from any funding agency in the public, commercial, or non-profit sectors.

\pdfbookmark[1]{Competing interests}{competinginterests}
\bmhead{Competing interests}
The author invented Indian patent IN346145, which relates to magnonic-crystal cavity architectures. The author declares no other competing interests.

\subsection*{Ethics approval and consent to participate}
Not applicable.

\section*{Consent for publication}
Not applicable.
\bmhead{Data availability}
The datasets generated and analysed during the current study are available from the corresponding author on reasonable request.

\bmhead{Materials availability}
Not applicable.

\bmhead{Code availability}
The MuMax3 input scripts, the QuTiP master-equation and sensitivity-projection code and the plane-wave band-structure code are available from the corresponding author on reasonable request.

\bmhead{Author contribution}
N.K. conceived the study, performed the micromagnetic, plane-wave, and master-equation simulations, analysed the results, and wrote the manuscript.

\bibliography{paper}

\end{document}

\title[Supplementary Information]{Supplementary Information for ``Magnonic cavity-enhanced quantum magnetometry with nitrogen-vacancy centre ensembles''}

\author*[1]{\fnm{Nikhil} \sur{Kumar}}\email{nikhilkumarcs@nitc.ac.in}

\affil*[1]{\orgdiv{Department of Electronics and Communication Engineering}, \orgname{National Institute of Technology Calicut}, \orgaddress{\city{Kozhikode}, \postcode{673601}, \state{Kerala}, \country{India}}}

\abstract{This Supplementary Information provides the static micromagnetic characterisation of the YIG/CoFeB magnonic crystal (Supplementary Note~1), the plane-wave method, the micromagnetic dispersion extraction and the defect-mode classification (Note~2), the derivation of the homodyne and analytic ODMR steady states (Note~3), the simulation parameters (Note~4) and the sensitivity projection (Note~5). Equations, figures and tables of the main text are cited as Eq.~(n), Fig.~n and Table~n; supplementary items carry the prefix S.}

\maketitle

%% =====================================================================
\section{Supplementary Note 1: Static relaxation and stray-field mapping}

\begin{figure}[htbp]
\centering
\includegraphics[width=\textwidth]{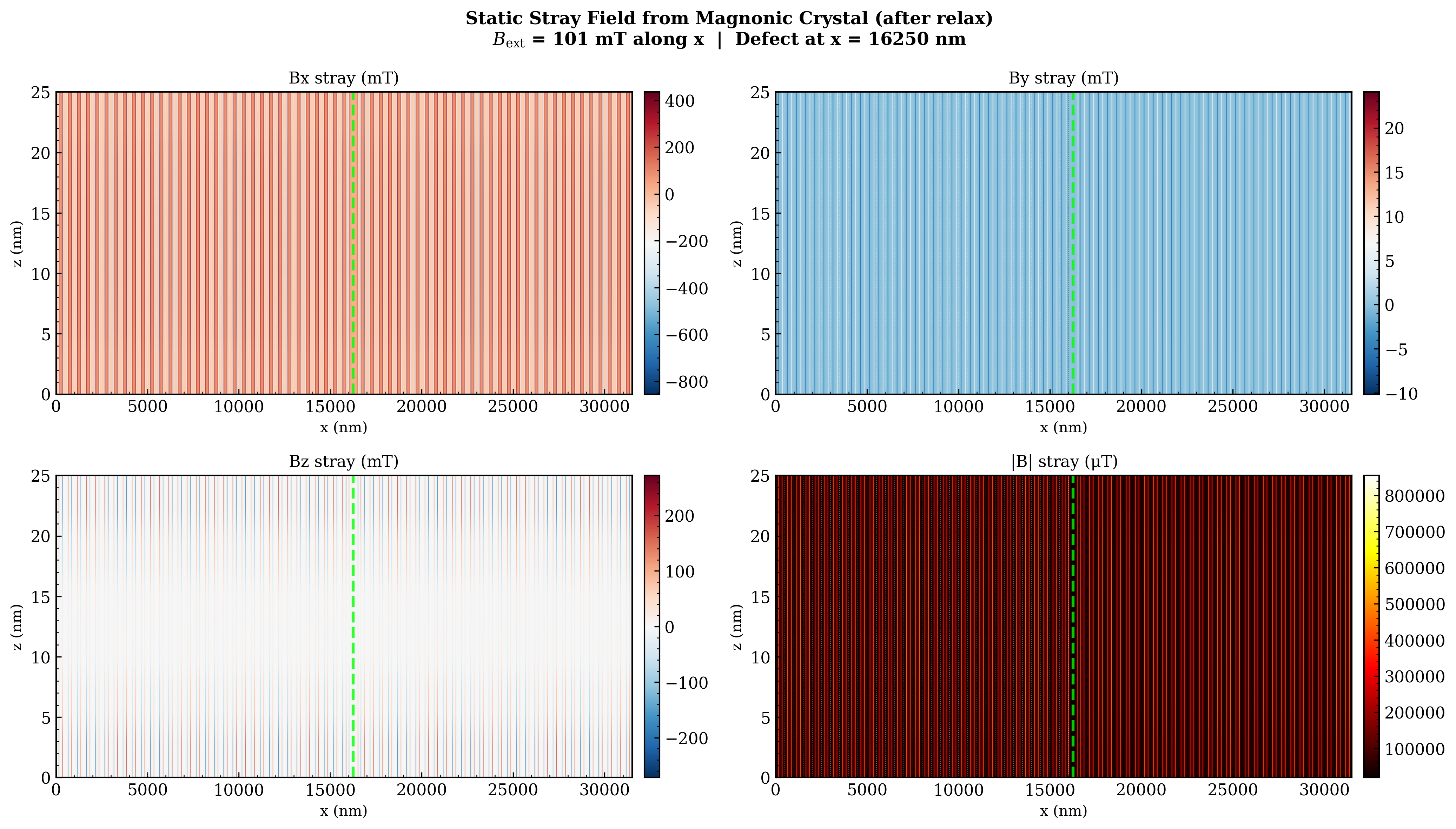}
\caption{\textbf{Static stray field of the relaxed magnonic crystal.} Maps of the static demagnetizing field components $B_{x}$, $B_{y}$ and $B_{z}$ (in mT) and of the magnitude $|B|$ (in $\mu$T) in the $x$--$z$ plane of the 1D YIG/CoFeB magnonic crystal after relaxation under $B_{\mathrm{ext}}=101.4$ mT applied along $x$. The green dashed line marks the defect position ($x=16{,}250$ nm).}\label{sfig:static-maps}
\end{figure}

\begin{figure}[htbp]
\centering
\includegraphics[width=\textwidth]{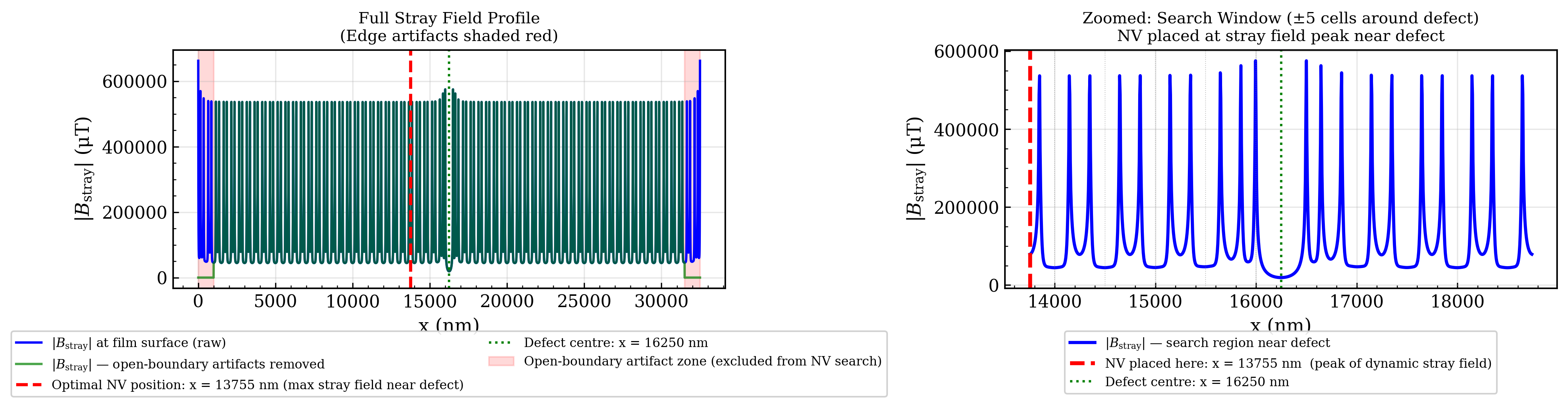}
\caption{\textbf{Static stray-field profile and choice of the NV position.} Left: full spatial profile of $|B_{\mathrm{stray}}|$ at the film surface; the red shaded zones near the ends are affected by open-boundary artefacts and are excluded from the analysis. Right: enlarged view of the search window around the defect, showing the selected NV position at $x=13{,}755$ nm (red dashed line) relative to the defect centre at $x=16{,}250$ nm (green dotted line).}\label{sfig:static-profile}
\end{figure}

The built-in MuMax3\cite{vansteenkiste2014design} relax() routine brings the device to a static energy minimum under a uniform bias field $B_{\mathrm{ext}}=101.4$ mT along $x$, representing the stray field of permanent magnets on the chip (main text, Fig.~1). This routine minimises the total micromagnetic energy (exchange, demagnetising, and Zeeman contributions) using an adaptive, torque-based steepest-descent scheme and stops when the maximum torque per cell falls below a preset threshold; at this stage, no time integration or excitation occurs. Supplementary Fig.~\ref{sfig:static-maps} maps the static demagnetising (stray) field components $B_{x}$, $B_{y}$ and $B_{z}$ of the relaxed configuration, and Supplementary Fig.~\ref{sfig:static-profile} plots the resulting field profile along the device. The static map confirms the crystal's periodic dipolar pattern, with a magnitude of several hundred millitesla at the film surface, and shows a local field dip at the defect centre, consistent with the locally modified magnetic landscape in the defect region.

The open boundaries of the simulated domain produce demagnetisation artefacts within about 1 $\mu$m of each end. The analysis excludes these zones (shaded in Supplementary Fig.~\ref{sfig:static-profile}) from all field and mode analysis and searches for the NV position within $\pm5$ unit cells of the defect.

%% =====================================================================
\section{Supplementary Note 2: Band-structure calculations and defect-mode classification}

\subsection{Plane-wave method}

\subsubsection{Plane-wave method details}
\paragraph{Geometry and assumptions.} The plane-wave method (PWM) used here follows the formulation developed for striped Co/permalloy magnonic waveguides in ref.~\cite{kumar2013spin}, which in turn adapts the plane-wave theory of three-dimensional magnonic crystals\cite{krawczyk2008plane,puszkarski2003magnonic,vasseur1996magnon} to a film of finite thickness. The pristine crystal is periodic along $x$ with lattice constant $a=s+d$, where $s$ and $d$ are the widths of the YIG and CoFeB segments. The film has thickness $t=25$ nm along $z$ and width $w$ along $y$. The bias field $\mathbf{H}_{0}=H_{0}\hat{\mathbf{x}}$, the direction of periodicity and the spin-wave vector are parallel, so the crystal operates in a backward volume geometry. The model assumes that the bias saturates both materials along $x$, neglects magnetocrystalline anisotropy in the small-signal dynamics and omits damping, which does not shift the eigenfrequencies to first order.

\paragraph{Equation of motion.} The magnetisation obeys the Landau--Lifshitz equation.
\begin{equation}
\frac{\partial\mathbf{M}}{\partial t}=-\gamma\mu_{0}\,\mathbf{M}\times\mathbf{H}_{\mathrm{eff}},\qquad
\mathbf{H}_{\mathrm{eff}}=H_{0}\hat{\mathbf{x}}+\mathbf{H}_{\mathrm{ex}}+\mathbf{H}_{\mathrm{d}},\label{seq:ll}
\end{equation}
where $\gamma$ is the gyromagnetic ratio and $\mu_{0}$ the vacuum permeability. The exchange field is
\begin{equation}
\mathbf{H}_{\mathrm{ex}}=\nabla\cdot\left(\lambda_{\mathrm{ex}}^{2}(x)\nabla\right)\mathbf{M},\qquad \lambda_{\mathrm{ex}}^{2}=\frac{2A_{\mathrm{ex}}}{\mu_{0}M_{s}^{2}},\label{seq:exchange}
\end{equation}
with exchange stiffness $A_{\mathrm{ex}}$ and saturation magnetisation $M_{s}$, both of which alternate between their YIG and CoFeB values. The magnetisation is the sum of a static part along the bias and a small-signal part,
\[
\mathbf{M}(\mathbf{r},t)=M_{s}(x)\left[\hat{\mathbf{x}}+\mathbf{m}(\mathbf{r},t)\right],\qquad \mathbf{m}=(0,m_{y},m_{z}),\qquad |\mathbf{m}|\ll1,
\]
where $\mathbf{m}$ is the magnetisation of the small-signal normalised to the local $M_{s}$. The demagnetising field splits in the same way into a static part and a dynamic part $\mathbf{H}_{\mathrm{d}}=\mathbf{H}(\mathbf{r})+\mathbf{h}(\mathbf{r},t)$, which in the magnetostatic approximation satisfy\cite{stancil2009spin}
\begin{equation}
\nabla\times\mathbf{H}=0,\quad \nabla\cdot\left(\mathbf{H}+M_{s}\hat{\mathbf{x}}\right)=0,\qquad \nabla\times\mathbf{h}=0,\quad \nabla\cdot\left(\mathbf{h}+M_{s}\mathbf{m}\right)=0.\label{seq:magnetostatics}
\end{equation}
For harmonic dynamics $\propto e^{i\omega t}$, keeping terms to first order in $\mathbf{m}$, the linearised equation of motion reads
\begin{equation}
i\omega\,m_{y}=-\gamma\mu_{0}\left[H_{i}(x)\,m_{z}-h_{z}^{\mathrm{tot}}\right],\qquad
i\omega\,m_{z}=\gamma\mu_{0}\left[H_{i}(x)\,m_{y}-h_{y}^{\mathrm{tot}}\right],\label{seq:linear}
\end{equation}
where $H_{i}(x)=H_{0}+H_{x}(x)$ is the static internal field, including the static demagnetising field $H_{x}$ produced by the magnetic charges at the YIG/CoFeB interfaces, and $\mathbf{h}^{\mathrm{tot}}$ is the sum of the dynamic exchange and dipolar fields.

\paragraph{Plane-wave expansion.} Every periodic material parameter is expanded over the reciprocal lattice vectors $G_{n}=2\pi n/a$. For the two-component step profile with the YIG segment centred at the origin,
\begin{equation}
M_{s}(G)=\begin{cases}M_{s}^{\mathrm{Y}}f+M_{s}^{\mathrm{C}}(1-f), & G=0,\\[2pt] \left(M_{s}^{\mathrm{Y}}-M_{s}^{\mathrm{C}}\right)f\,\dfrac{\sin(Gs/2)}{Gs/2}, & G\neq0,\end{cases}\qquad f=\frac{s}{a},\label{seq:fourier}
\end{equation}
and $\lambda_{\mathrm{ex}}^{2}(G)$ follows in the same way, where the superscripts Y and C denote YIG and CoFeB. By Bloch's theorem, the small-signal magnetisation takes the form of
\begin{equation}
m_{y,z}(\mathbf{r})=\sum_{G}m_{y,z}(G)\,e^{i(k+G)x}\,e^{ik_{y}y},\label{seq:bloch}
\end{equation}
with the Bloch wave vector $k$ in the first Brillouin zone, $-\pi/a\leq k\leq\pi/a$. Across the width, effective dipolar pinning at the lateral edges quantises the transverse wave vector as $k_{y}=r\pi/w$, with $r$ an integer that depends on the type of pinning\cite{guslienko2002effective,kumar2013spin}; the bands in Fig.~2a of the main text use the lowest-width mode. Across the 25 nm thickness, we take the dynamic profile as uniform, which is accurate for the lowest-thickness mode of a thin film compared with the exchange and dipolar lengths that matter at these frequencies\cite{kalinikos1986theory}.

\paragraph{Static and dynamic dipolar fields.} The magnetostatic potential of the periodic magnetisation follows from Eq.~\eqref{seq:magnetostatics} with continuity of the potential and of the normal induction at the film surfaces\cite{kaczer1974demagnetizing,kumar2013spin}. Inside the film, the static field along $x$ has the profile $H_{x}(x,z)=-\sum_{G\neq0}M_{s}(G)\left[1-\cosh(|G|z)\,e^{-|G|t/2}\right]e^{iGx}$, whose average thickness gives the following.
\begin{equation}
H_{x}(G)=-M_{s}(G)\left[1-S(|G|)\right]\quad(G\neq0),\qquad H_{x}(0)=0,\qquad S(q)=\frac{1-e^{-qt}}{qt}.\label{seq:static}
\end{equation}
This field is largest in magnitude in the low-$M_{s}$ YIG segments next to the CoFeB stripes, so that the internal field $H_{i}(x)$ is strongly non-uniform. For the dynamic field, each Bloch component with in-plane wave vector $q_{n}=\sqrt{(k+G_{n})^{2}+k_{y}^{2}}$ produces, after averaging over the thickness,
\begin{equation}
h_{z}(G_{n})=-S(q_{n})\,\mathcal{M}_{z}(G_{n}),\qquad h_{y}(G_{n})=-\left[1-S(q_{n})\right]\frac{k_{y}^{2}}{q_{n}^{2}}\,\mathcal{M}_{y}(G_{n}),\label{seq:dynamic}
\end{equation}
where $\mathcal{M}_{y,z}(G_{n})=\sum_{j}M_{s}(G_{n}-G_{j})\,m_{y,z}(G_{j})$ are the Fourier coefficients of dynamic magnetisation $M_{s}(x)\,m_{y,z}(x)$. The component outside the plane $h_{z}$ is the thin-film depolarising field, which tends to $-M_{s}m_{z}$ for $q_{n}t\to0$, while the component inside the plane $h_{y}$ vanishes for a uniform mode across the width ($k_{y}=0$).

\paragraph{Eigenvalue problem.} Substituting Eqs.~\eqref{seq:exchange}, \eqref{seq:fourier}--\eqref{seq:dynamic} into Eq.~\eqref{seq:linear} and collecting the coefficient of each plane wave turns the spin-wave problem into the matrix eigenvalue problem\cite{krawczyk2008plane,kumar2013spin}
\begin{equation}
\hat{M}\,\mathbf{m}_{k}=i\omega\,\mathbf{m}_{k},\qquad
\mathbf{m}_{k}=\left[m_{y}(G_{1}),\dots,m_{y}(G_{N}),\,m_{z}(G_{1}),\dots,m_{z}(G_{N})\right]^{T},\qquad
\hat{M}=\begin{pmatrix}0 & \hat{M}^{yz}\\ \hat{M}^{zy} & 0\end{pmatrix},\label{seq:eigen}
\end{equation}
with the sub-matrices;
\begin{align}
M^{yz}_{ij}&=-\gamma\mu_{0}\left[H_{i}(G_{i}-G_{j})+E_{ij}+S(q_{i})\,M_{s}(G_{i}-G_{j})\right],\nonumber\\
M^{zy}_{ij}&=\gamma\mu_{0}\left[H_{i}(G_{i}-G_{j})+E_{ij}+\left[1-S(q_{i})\right]\frac{k_{y}^{2}}{q_{i}^{2}}\,M_{s}(G_{i}-G_{j})\right],\label{seq:blocks}\\
E_{ij}&=\sum_{l}\left[(k+G_{i})(k+G_{l})+k_{y}^{2}\right]\lambda_{\mathrm{ex}}^{2}(G_{i}-G_{l})\,M_{s}(G_{l}-G_{j}),\nonumber
\end{align}
where $H_{i}(G)=H_{0}\,\delta_{G,0}+H_{x}(G)$. The exchange term $E_{ij}$ is the Fourier representation of Eq.~\eqref{seq:exchange}; because it is written for the full magnetisation $M_{s}(x)\left[\hat{\mathbf{x}}+\mathbf{m}\right]$, it carries the exchange coupling across the YIG/CoFeB interfaces without separate boundary conditions. For a uniform film ($N=1$, $k_{y}=0$), Eq.~\eqref{seq:eigen} reduces to the backward-volume dispersion $\omega^{2}=\gamma^{2}\mu_{0}^{2}\left(H_{0}+\lambda_{\mathrm{ex}}^{2}M_{s}k^{2}\right)\left(H_{0}+\lambda_{\mathrm{ex}}^{2}M_{s}k^{2}+S(k)M_{s}\right)$, which tends to the Kittel frequency for $k\to0$ and decreases with $k$\cite{kalinikos1986theory,stancil2009spin}, a useful cheque of implementation.

\paragraph{Numerical procedure.} The calculation proceeds as follows: (i) build the Fourier coefficients of $M_{s}$ and $\lambda_{\mathrm{ex}}^{2}$ from Eq.~\eqref{seq:fourier} for $|n|\leq n_{\max}$, so that $N=2n_{\max}+1$; (ii) compute the static internal field from Eq.~\eqref{seq:static}; (iii) for each $k$ on a grid across the first Brillouin zone, assemble $\hat{M}$ from Eq.~\eqref{seq:blocks} and diagonalise it; (iv) keep the eigenvalues with positive frequency, $f_{n}(k)=\omega_{n}(k)/2\pi$, which form the bands, and use the eigenvectors to reconstruct the spatial mode profiles through Eq.~\eqref{seq:bloch}. Because the step profiles of Eq.~\eqref{seq:fourier} have Fourier coefficients that decay only as $1/n$, and because the contrast in $M_{s}$ between CoFeB and YIG is large, $n_{\max}$ increases until the bands below 4 GHz change by less than 1\%; in ref.~\cite{kumar2013spin} about 30 reciprocal lattice vectors were sufficient for the bands, and the static field needed more harmonics, converging as a damped oscillation in the number of vectors. The present calculation uses $n_{\max}=$ 30. The calculation uses only the geometry and material parameters of the micromagnetic model (Supplementary Table~\ref{stab:params}), without fitting to the micromagnetic output, and applies only to the strictly periodic crystal.

\paragraph{Physical content of the bands.} Two features of the PWM bands in Fig.~2a of the main text follow directly from this formulation. First, the periodic modulation of $M_{s}$ and $A_{\mathrm{ex}}$ couples plane waves that differ by a reciprocal lattice vector, and Bragg scattering opens gaps at the centre and boundary of the zone; the gap widths grow with the contrast between the two materials and depend on the filling fraction $f$, as shown for the Co/permalloy waveguide in ref.~\cite{kumar2013spin}. Second, the strongly non-uniform static internal field of Eq.~\eqref{seq:static} makes low-frequency spin waves concentrate their amplitude in the regions of low internal field, where they form standing waves with nearly zero group velocity. They appear as nearly flat bands below about 1.5 GHz in Fig.~2a, the same mechanism that produces the flat low-frequency bands of the striped waveguide\cite{kumar2013spin}. At higher frequencies, the bands become dispersive and the gap between 2.5 and 3.0 GHz opens, hosting the defect mode. This combined PWM and micromagnetic strategy follows the approach used previously for striped waveguides, magnonic rings, and antidot crystals with a ring defect \cite{kumar2013spin,kumar2014iterative,venkat2014micromagnetic}.

\subsection{Micromagnetic dispersion extraction}

To obtain a dispersion relation rather than a single-point spectrum, a broadband pulse that is flat in frequency up to its cut-off excites the relaxed crystal, driving all spin-wave modes in that band simultaneously and with comparable amplitude. The solver records the magnetisation in every cell along the direction of periodicity, and a two-dimensional Fourier transform, first from space to wave vector and then from time to frequency, converts the $m(x,t)$ record into the dynamical structure factor $S(k,\omega)$, whose intensity ridges trace the dispersion without assuming any band model. Beyond the bands reproduced by the PWM (main text, Fig.~2a), the micromagnetic map shows a few sharp features near 2.4 and 2.55 GHz and a diffuse continuum at about 1.8--2.3 GHz. The linear, single-mode PWM does not capture these features. They are consistent with higher-order width or thickness modes, which lie outside the lowest-mode ansatz of Eq.~\eqref{seq:bloch}, or with nonlinear spin-wave processes.

\subsection{Defect-supercell analysis}

\begin{figure}[htbp]
\centering
\includegraphics[width=0.6\textwidth]{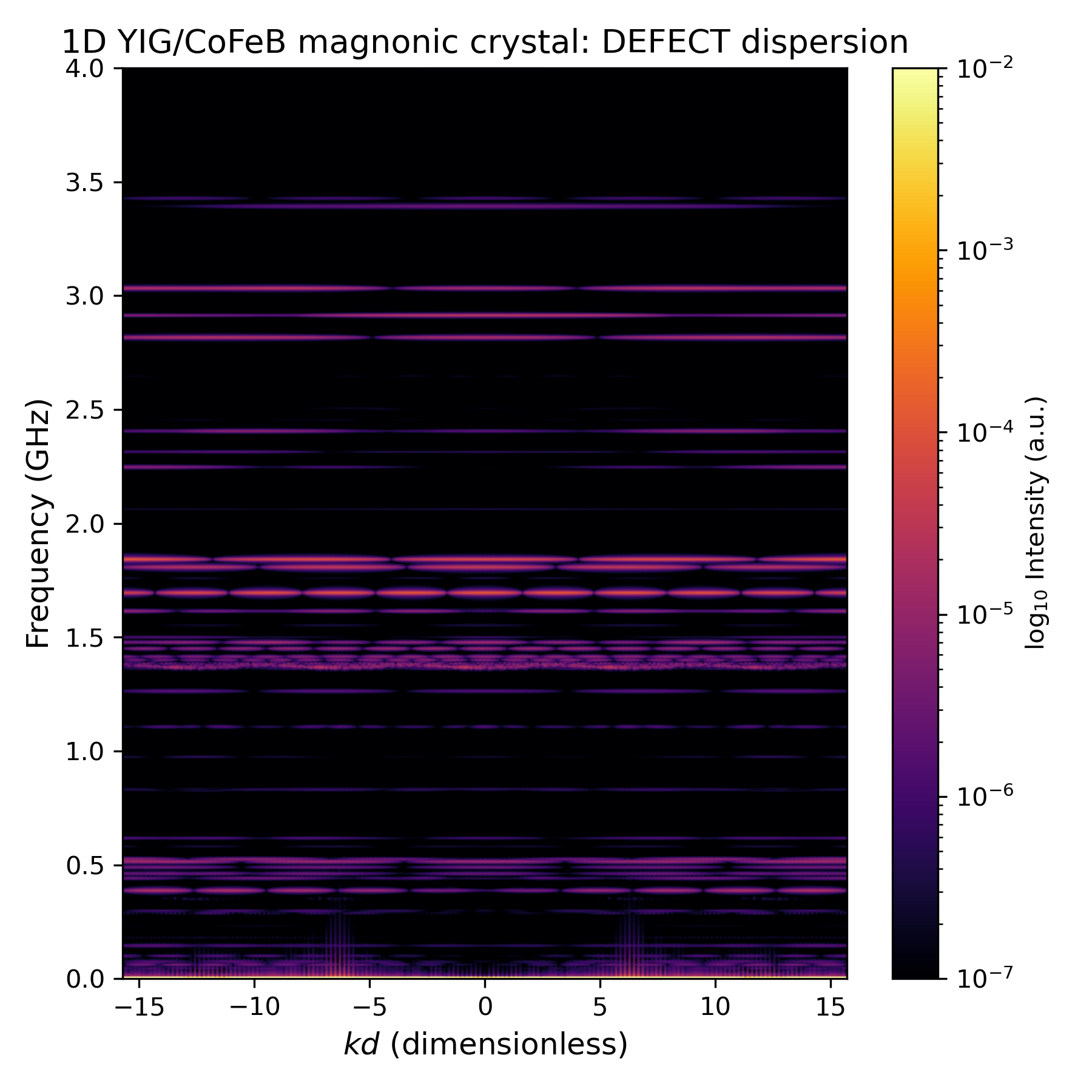}
\caption{\textbf{Dispersion of the crystal containing a point defect.} Dispersion of the 1D YIG/CoFeB crystal with a single point defect, computed over the whole simulated domain with the defect-supercell method. Breaking the strict periodicity folds the bulk bands into a dense manifold of closely spaced, nearly flat replica branches (most prominent between 1.3 and 1.9 GHz) separated by about $2\pi/L$ for a supercell of length $L$. A mode is classified as defect-localised if its frequency lies within a bulk gap of the pristine dispersion (main text, Fig.~2a) and its real-space profile is confined to the defect.}\label{sfig:defect}
\end{figure}

A single point defect removes the discrete translational symmetry assumed in the single-unit-cell PWM calculation. The analysis therefore adopts the standard defect-supercell approach, which treats the entire simulated domain, including the bulk crystal and the defect, as one large effective unit cell.\cite{venkat2014micromagnetic,kumar2022enhanced} This folds the bulk bands into a dense set of closely spaced, nearly flat replica branches separated by about $2\pi/L$ in wave vector for a supercell of length $L$ (Supplementary Fig.~\ref{sfig:defect}). Two independent criteria distinguish a genuine defect-localised mode from the folded bulk replicas.\cite{kruglyak2006spin,gallardo2018symmetry,zhang2016spin} First, the mode frequency must lie inside a gap of the pristine band structure rather than on a folded bulk band. Second, its spatial profile must be exponentially confined to the defect rather than extended across the crystal, as the per-cell mode profile in Fig.~4a of the main text shows.

\section{Supplementary Note 3: Homodyne and analytic ODMR steady states}

Both start from the Tavis--Cummings Hamiltonian of Eq.~(1), in which each $^{14}$N projection $m_{I}\in\{-1,0,+1\}$ forms an independent two-level system on the $m_{s}=0\leftrightarrow m_{s}=+1$ transition, coupled with strength $g_{\mathrm{ens}}$ to the defect magnon mode $a$ of the loss rate $\gamma_{0}=\kappa_{m}$. They differ in what drives the system and in what is measured. In the homodyne route, a weak probe drives the mode and the reflected microwave is detected, so the spins stay close to their polarised ground state; in the ODMR route, a stronger drive reaches the spins through the mode, and the population transferred out of $m_{s}=0$ is detected optically.

\subsection{Homodyne response}

For the homodyne readout, a weak coherent probe of amplitude $\epsilon$ and frequency $\omega_{p}$ is coupled into the mode through the antenna. In the frame rotating at $\omega_{p}$, the Hamiltonian reads\cite{kumar2026cavity}
\begin{equation}
H_{\mathrm{p}}=\Delta_{c}\,a^{\dagger}a+\sum_{j=-1}^{+1}\left[\frac{\Delta_{j}}{2}\,\sigma_{j}^{z}+g_{\mathrm{ens}}\left(a\,\sigma_{j}^{+}+a^{\dagger}\sigma_{j}^{-}\right)\right]-\epsilon\left(a+a^{\dagger}\right),\label{seq:probe}
\end{equation}
where $\Delta_{c}=\omega_{c}-\omega_{p}$ is the detuning of the mode from the probe and $\Delta_{j}=\omega_{j}-\omega_{p}=\Delta_{s}+m_{I}A_{\mathrm{hf}}+\Delta_{c}$ is the detuning of sector $j$, with $\Delta_{s}=\omega_{s}-\omega_{c}$ set by the bias field. The probe frequency is the swept variable, so $\Delta_{c}$ and all three $\Delta_{j}$ change together while the spin frequencies themselves stay fixed; omitting $\Delta_{c}$ from $\Delta_{j}$ would make the spins follow the probe and produce spurious features. Treating the mode and the spin coherences as amplitudes driven by the probe and by vacuum or thermal noise inputs $a^{\mathrm{in}}$ and $b_{j}^{\mathrm{in}}$ gives the quantum Langevin equations \cite{gardiner1985input}.
\begin{align}
\frac{da}{dt} & =-\left(i\Delta_{c}+\frac{\gamma_{0}}{2}\right)a-i\,g_{\mathrm{ens}}\sum_{j}\sigma_{j}^{-}+i\epsilon+\sqrt{\gamma_{0}}\,a^{\mathrm{in}},\label{seq:langevin-a}\\
\frac{d\sigma_{j}^{-}}{dt} & =-\left(i\Delta_{j}+\frac{\Gamma}{2}\right)\sigma_{j}^{-}+i\,g_{\mathrm{ens}}P\,a+\sqrt{\Gamma}\,b_{j}^{\mathrm{in}},\label{seq:langevin-s}
\end{align}
in which $P=\langle\sigma_{j}^{z}\rangle\simeq-1$ is the spin polarisation maintained by optical pumping and $\Gamma=2/T_{2}^{*}$ is the full spin linewidth, including relaxation, dephasing, and inhomogeneous broadening of the ensemble.

In steady state, the time derivatives vanish, and Eq.~\eqref{seq:langevin-s} gives each spin coherence in terms of the mode amplitude, $\sigma_{j}^{-}=i\,g_{\mathrm{ens}}P\,a/(i\Delta_{j}+\Gamma/2)$. Physically, every hyperfine sector behaves as a small resonator that is driven by the mode and reacts back on it. Substituting this result into Eq.~\eqref{seq:langevin-a} and taking the mean, in which the noise inputs vanish, gives the closed form of Eq.~(5) of the main text. The spins enter only through the term $g_{\mathrm{ens}}^{2}P\sum_{j}(i\Delta_{j}+\Gamma/2)^{-1}$, whose imaginary part pulls the mode frequency and whose real part adds loss; each pull changes sign as a hyperfine line crosses the mode, which produces the three steep edges of Fig.~6c. The probe amplitude coupled into the mode is $\epsilon=\sqrt{\kappa_{\mathrm{ext}}}\,a^{\mathrm{mw}}$, where $a^{\mathrm{mw}}$ is the incident field, and $\kappa_{\mathrm{ext}}$ is the antenna--mode coupling rate. By input--output theory the reflected field is $a^{\mathrm{out}}=a^{\mathrm{mw}}-\sqrt{\kappa_{\mathrm{ext}}}\,\alpha$, so the reflection coefficient of the antenna is $r=1-\kappa_{\mathrm{ext}}\,\alpha/\epsilon$, the quantity used for the sensitivity projection of the Supplementary Note~5. In the weak-probe limit, the spins remain polarised and $P$ stays at $-1$, so the closed form is exact for the model; the point-by-point agreement with the master equation in Fig.~6a,b ($R^{2}>0.9999$) confirms this and shows that the slopes entering the sensitivity are free of numerical artefacts.

\subsection{Analytic ODMR spectrum}

In the ODMR route, the antenna drives the defect mode, and the transverse field of the mode drives each hyperfine sector with the Rabi frequency $\Omega(\Delta_{d})$ of Eq.~(3) of the main text and a detuning $\delta_{j}=\Delta_{s}+m_{I}A_{\mathrm{hf}}-\Delta_{d}$ from the drive. The mode also couples the spins to its own loss channel. When $g_{\mathrm{ens}}$ is small compared to $\gamma_{0}/2$, the mode responds faster than the spins and can be eliminated adiabatically, leaving each sector with two additional terms, a Purcell decay rate and a mode-induced frequency pull,
\begin{equation}
\Gamma_{\mathrm{P},j}=\frac{g_{\mathrm{ens}}^{2}\,\gamma_{0}}{\Delta_{sj}^{2}+\gamma_{0}^{2}/4},\qquad
\delta_{\mathrm{L},j}=\frac{g_{\mathrm{ens}}^{2}\,\Delta_{sj}}{\Delta_{sj}^{2}+\gamma_{0}^{2}/4},\qquad \Delta_{sj}=\Delta_{s}+m_{I}A_{\mathrm{hf}}.\label{seq:purcell}
\end{equation}
The Purcell decay is largest when the sector is resonant with the mode ($\Delta_{sj}\approx0$) and broadens the line; the pull vanishes at resonance, is largest at $|\Delta_{sj}|\approx\gamma_{0}/2$ and pushes the line away from the mode frequency. Each sector is then an ordinary driven two-level system, and the steady state of its Bloch equations gives the population of $|m_{s}=+1\rangle$,
\begin{equation}
P_{1,j}=\frac{1}{2}\,\frac{X_{j}}{\left(\delta_{j}+\delta_{\mathrm{L},j}\right)^{2}+G_{2,j}^{2}+X_{j}},\qquad X_{j}=\Omega^{2}\frac{G_{2,j}}{G_{1,j}},\label{seq:bloch-ss}
\end{equation}
with the population decay rate $G_{1,j}=T_{1}^{-1}+\Gamma_{p}+\Gamma_{\mathrm{P},j}$, the coherence decay rate $G_{2,j}=G_{1,j}/2+\gamma_{\phi}$ and the pure dephasing rate $\gamma_{\phi}=1/T_{2}^{*}-1/(2T_{1})$. Equation~\eqref{seq:bloch-ss} contains saturation through $X_{j}$, which increases as $\Omega^{2}$ and limits $P_{1,j}$ to $\frac{1}{2}$, and power broadening, which widens the line as $\Omega$ increases. The ODMR signal of Fig.~7 is the average $\frac{1}{3}\sum_{j}P_{1,j}$ with equal weights, normalised once over the whole map.

This analytic spectrum has a clear range of validity. Without the mode ($g_{\mathrm{ens}}=0$), the Purcell and pull terms vanish, and Eq.~\eqref{seq:bloch-ss} reproduces the master-equation solution exactly. With the mode, it is approximate. For $g_{\mathrm{ens}}/2\pi=0.25$ MHz and $\gamma_{0}/2\pi=1.0$ MHz, it reproduces the peak positions to within 0.012 MHz r.m.s. and the overall pattern of the simulated map ($R^{2}=0.91$), but it overestimates the heights of the outer hyperfine lines near the mode line (0.41 instead of 0.30 at $\Delta_{s}=0$). The adiabatic elimination neglects the coherent back-action of the mode on the driven spins and the mode-mediated coupling between the three sectors, both of which matter when $g_{\mathrm{ens}}$ is not small compared to $\gamma_{0}/2$. The deviation decreases as $g_{\mathrm{ens}}$ decreases; therefore, the analytic theory fixes the positions and overall structure of the spectrum, while the full master equation determines the quantitative peak heights.

\section{Supplementary Note 4: Simulation parameters}

Supplementary Table~\ref{stab:params} lists the parameters used in the micromagnetic, plane-wave, and master-equation simulations. The NV--magnon coupling and the defect-mode linewidth are representative values; their influence on the sensitivity, and that of the antenna coupling, is mapped in Fig.~8 of the main text.

\begin{table}[htbp]
\caption{Simulation parameters.}\label{stab:params}
\begin{tabular}{@{}lll@{}}
\toprule
Parameter & Value & Notes\\
\midrule
YIG stripe width, $s$ & 300 nm & \\
CoFeB stripe width, $d$ & 200 nm & $a=s+d$\\
Film thickness, $t$ & 25 nm & \\
Cell size, $\Delta x\times\Delta y\times\Delta z$ & $5\times 5 \times 5 \times \text{nm}^{3}$ & \\
Simulated length & $\approx32{,}000$ nm & open boundaries\\
Defect position & 16,250 nm & \\
YIG: $M_{s}$, $A_{\mathrm{ex}}$, $\alpha$ & $140$ kA\,m$^{-1}$, $3.7$ pJ\,m$^{-1}$, $1\times10^{-4}$ & YIG refs.~\cite{klingler2015measurements,chumak2015magnon}\\
CoFeB: $M_{s}$, $A_{\mathrm{ex}}$, $\alpha$ & $1250$ kA\,m$^{-1}$, $15$ pJ\,m$^{-1}$, $4.2\times10^{-3}$ & Co$_{40}$Fe$_{40}$B$_{20}$, ref.~\cite{conca2013low}\\
Bias field, $B_{\mathrm{ext}}$ & 101.4 mT & along $x$ (magnet stray field)\\
Excitation & sinc pulse, 10 mT, 5 GHz cut-off & uniform\\
PWM harmonics, $n_{\max}$ & 30 & $N=2n_{\max}+1=61$\\
Defect-mode frequency, $f_{\mathrm{mode}}$ & 2.910 GHz & MuMax3\\
Field enhancement, $\mathcal{E}=B_{\mathrm{local}}/B_{\mathrm{signal}}$ & 0.3125 & CW, $B_{\mathrm{signal}}=0.5$ mT\\
NV zero-field splitting, $f_{\mathrm{ZFS}}$ & 2.87 GHz & \\
Hyperfine splitting, $A_{\mathrm{hf}}$ & 2.16 MHz & equal $m_{I}$ weights\\
Rabi frequency on the mode resonance, $\Omega_{0}/2\pi$ & 0.10 MHz & Eq.~(3) of the main text\\
Defect-mode linewidth, $\kappa_{m}/2\pi=\gamma_{0}/2\pi$ & 1.0 MHz & representative\\
NV--magnon coupling, $g_{\mathrm{ens}}/2\pi$ & 0.25 MHz & representative\\
Ensemble dephasing time, $T_{2}^{*}$ & 0.5 $\mu$s & $\Gamma/2\pi=0.64$ MHz\\
Longitudinal relaxation time, $T_{1}$ & 1 ms & \\
Optical repolarisation rate, $\Gamma_{p}$ & 0.1 $\mu$s$^{-1}$ & \\
Homodyne probe amplitude, $\epsilon/2\pi$ & 0.01 MHz & linear response\\
\botrule
\end{tabular}
\end{table}

%% =====================================================================
\section{Supplementary Note 5: Sensitivity projection}

\subsection{Thermal limit}

At the operating point $\Delta_{c}=0$, the antenna reflects the probe with coefficient $r(\omega_{s})=1-\kappa_{\mathrm{ext}}\,\alpha/\epsilon$ (Supplementary Note~3). A field change $\delta B$ shifts the spin frequency by $\gamma_{e}\delta B$ and moves the dispersive quadrature $X=\mathrm{Im}\,r$ of the reflected wave by $\sqrt{P_{\mathrm{in}}}\,|dX/d\omega_{s}|\,\gamma_{e}\delta B$, in units of $\sqrt{\mathrm{W}}$, for an input power $P_{\mathrm{in}}$. The thermal (Johnson--Nyquist) and vacuum noise in one quadrature has the one-sided spectral density $S_{b}=(k_{B}T+hf/2)/2$\cite{clerk2010introduction}. The minimum detectable field per unit bandwidth on edge $j$ is therefore
\begin{equation}
\eta_{j}^{\mathrm{th}}=\frac{\sqrt{3}\,\sqrt{S_{b}}}{\gamma_{e}\sqrt{P_{\mathrm{in}}}\,\left|dX/d\omega_{s}\right|_{\max,j}},\label{seq:eta-th}
\end{equation}
and the edges combine through Eq.~(8) of the main text. For an empty resonator with internal loss rate $\kappa_{\mathrm{int}}$, the reflected signal is largest at critical coupling, $\kappa_{\mathrm{ext}}=\kappa_{\mathrm{int}}$, and degrades by $(\kappa_{\mathrm{int}}+\kappa_{\mathrm{ext}})^{2}/(4\kappa_{\mathrm{int}}\kappa_{\mathrm{ext}})$ away from it\cite{pozar2012microwave}. With the spins present, the mode loses additional energy to the spin ensemble. Near a steep edge, where the operating spin feature lies within about $\Gamma/2$ of the mode, Eq.~(5) of the main text adds a loss rate of order $4g_{\mathrm{ens}}^{2}/\Gamma$, so that the effective internal loss becomes $\kappa_{\mathrm{int}}+4g_{\mathrm{ens}}^{2}/\Gamma$ and the optimum antenna coupling moves to
\begin{equation}
\kappa_{\mathrm{ext}}^{\mathrm{opt}}\approx\kappa_{\mathrm{int}}+\frac{4g_{\mathrm{ens}}^{2}}{\Gamma}.\label{seq:kopt}
\end{equation}
The complete calculation of Fig.~8b, in which the slopes are recomputed from Eq.~(5) with $\kappa_{m}=\kappa_{\mathrm{int}}+\kappa_{\mathrm{ext}}$ for each $\kappa_{\mathrm{ext}}$, places the optima at $\kappa_{\mathrm{ext}}/\kappa_{\mathrm{int}}=1.78$ and 3.98 for $\Gamma/2\pi=0.64$ and 0.17 MHz, in agreement with Eq.~\eqref{seq:kopt} (1.79 and 3.98). Weak coupling is penalised more strongly than for an empty resonator: at $\kappa_{\mathrm{ext}}=0.05\,\kappa_{\mathrm{int}}$ the sensitivity degrades by a factor of 8 and 13 relative to critical coupling, compared to 5.5 of the empty resonator formula.

\subsection{Technical-noise excess}

A practical homodyne chain (low-noise amplifier, mixer, local oscillator and data acquisition) adds noise above the thermal limit. Its size is estimated from ref.~\cite{kumar2026cavity}: Eq.~\eqref{seq:eta-th} evaluated with the parameters of that work ($-40$ dBm probe, $g_{\mathrm{ens}}/2\pi=0.26$ MHz, $\Gamma/2\pi=167$ kHz, nine features, and an assumed cavity linewidth of 0.25 MHz) gives 1.1 pT$/\sqrt{\mathrm{Hz}}$, whereas the measured sensitivity is 12 pT$/\sqrt{\mathrm{Hz}}$, a technical-noise excess of 10.9. The projection for the magnonic device applies the same excess, which assumes a readout chain of comparable quality.

\subsection{Frequency dependence of the sensitivity}

To obtain the sensitivity as a function of the frequency $f$ of the field to be measured (Fig.~6d of the main text), the field is taken to oscillate, $\delta B(t)=\delta B\cos(2\pi ft)$, so that each spin detuning is modulated, $\Delta_{j}\to\Delta_{j}+\gamma_{e}\delta B\cos(2\pi ft)$. Writing the mode and spin amplitudes as their steady-state values at the operating point ($\Delta_{c}=0$, $\Delta_{s}$ on the steepest edge) plus small deviations $\mathbf{x}=(\delta a,\delta\sigma_{-1}^{-},\delta\sigma_{0}^{-},\delta\sigma_{+1}^{-})$, Eqs.~\eqref{seq:langevin-a} and \eqref{seq:langevin-s} give, to first order,
\begin{equation}
\frac{d\mathbf{x}}{dt}=-\hat{A}\,\mathbf{x}+\mathbf{b}\,\gamma_{e}\delta B(t),\qquad b_{j}=-i\,\bar{\sigma}_{j}^{-},\label{seq:linearised}
\end{equation}
where $\hat{A}$ contains the mode and spin decay rates and detunings and the couplings $\pm i g_{\mathrm{ens}}$, and $\bar{\sigma}_{j}^{-}$ is the steady-state coherence of sector $j$. For a modulation $e^{-i2\pi ft}$, the solution is $\mathbf{x}=(\hat{A}-i2\pi f)^{-1}\mathbf{b}$. The response of the dispersive quadrature to the real field is therefore
\begin{equation}
Y_{j}(f)=\frac{H_{j}(f)-H_{j}^{*}(-f)}{2i},\qquad H_{j}(f)=\left[(\hat{A}-i2\pi f)^{-1}\mathbf{b}\right]_{a},\label{seq:transfer}
\end{equation}
which reduces to the static slope $dX/d\omega_{s}$ as $f\to0$. The output noise of one quadrature is the thermal and vacuum noise of the room-temperature line, $S_{b}=(k_{B}T+hf_{c}/2)/2$, which is white at the modulation frequencies considered because they are far below the 2.9 GHz carrier. The sensitivity of edge $j$ is then $\eta_{j}(f)=\eta_{j}^{\mathrm{th}}|Y_{j}(0)|/|Y_{j}(f)|$, multiplied by the technical-noise excess for the projected curve, and the edges combine through Eq.~(8) of the main text. The response is flat while $f$ is small compared with the linewidths of the coupled system, so the sensitivity is constant up to about 100 kHz, and it falls above a $-3$ dB bandwidth of about 360 kHz, set by the spin linewidth $\Gamma$ and the mode linewidth $\kappa_{m}$. Near 2 MHz the modulation sidebands of one hyperfine line reach its neighbours, separated by $A_{\mathrm{hf}}=2.16$ MHz, which produces small features in $\eta(f)$. Technical $1/f$ noise and pickup at low frequencies, present in any real readout, are not part of the model.

\begin{table}[htbp]
\caption{Inputs and results of the sensitivity projection (critical coupling, $P_{\mathrm{in}}=-40$ dBm, $T=300$ K).}\label{stab:sens}
\begin{tabular}{@{}lll@{}}
\toprule
Quantity & Value & Notes\\
\midrule
Edge positions, $\Delta_{s}/2\pi$ & $-2.18$, 0, $+2.18$ MHz & Fig.~6c of the main text\\
Thermal limit, single edge & 14.7 pT$/\sqrt{\mathrm{Hz}}$ & Eq.~\eqref{seq:eta-th}\\
Thermal limit, three edges & 8.5 pT$/\sqrt{\mathrm{Hz}}$ & Eq.~(8), $M=3$\\
Technical-noise excess & 10.9 & from ref.~\cite{kumar2026cavity}\\
Projected best single edge & 161 pT$/\sqrt{\mathrm{Hz}}$ & \\
Projected three edges & 93 pT$/\sqrt{\mathrm{Hz}}$ & multiplexing gain 1.72\\
Sensitivity bandwidth ($-3$ dB) & $\approx360$ kHz & Fig.~6d\\
Optimum for $\Gamma/2\pi=0.64$ MHz & 72 pT$/\sqrt{\mathrm{Hz}}$ at $g_{\mathrm{ens}}/2\pi\approx0.44$ MHz & Fig.~8a\\
Optimum for $\Gamma/2\pi=0.17$ MHz & 19 pT$/\sqrt{\mathrm{Hz}}$ at $g_{\mathrm{ens}}/2\pi\approx0.20$ MHz & Fig.~8a\\
Optimum antenna coupling, $\Gamma/2\pi=0.64$ MHz & $\kappa_{\mathrm{ext}}=1.8\,\kappa_{\mathrm{int}}$, 85 pT$/\sqrt{\mathrm{Hz}}$ & Fig.~8b\\
Optimum antenna coupling, $\Gamma/2\pi=0.17$ MHz & $\kappa_{\mathrm{ext}}=4.0\,\kappa_{\mathrm{int}}$, 13 pT$/\sqrt{\mathrm{Hz}}$ & Fig.~8b\\
\botrule
\end{tabular}
\end{table}

\bibliography{paper}